\documentclass[fleqn,usenatbib]{mnras}
\usepackage{aas_macros}
\usepackage{abbreviations}
\usepackage{amssymb}
\usepackage{bigdelim}
\usepackage{bm}
\usepackage{cprotect}
\usepackage{etoolbox}
\usepackage{gensymb}
\usepackage{graphicx}
\usepackage{hyperref}
\usepackage{mathtools}
\usepackage{multirow}
\usepackage{multicol}
\usepackage{subcaption}
\usepackage{threeparttable}
\usepackage{upgreek}
\usepackage{xcolor}
\usepackage{newtxtext,newtxmath}
\usepackage[T1]{fontenc}
\usepackage[yyyymmdd]{datetime}

\def\xmm{\textit{XMM-Newton}}

\hypersetup{
  colorlinks   = true, 
  urlcolor     = blue, 
  linkcolor    = red, 
  citecolor   = blue 
}

\title[Abell 3266: fossils, relics, and haloes]{Radio fossils, relics, and haloes in Abell 3266: cluster archaeology with ASKAP-EMU and the ATCA}

\author[C.~J.~Riseley et al.]{C.~J.~Riseley$^{1,2,3}$\thanks{Corresponding author email: \url{christopher.riseley@unibo.it}}, \hspace{0.001cm} E.~Bonnassieux$^{1,2}$, T.~Vernstrom$^{4,3}$, T.~J.~Galvin$^{5,3}$, A.~Chokshi$^{6,7,8}$, A.~Botteon$^{9}$,
\newauthor
K.~Rajpurohit$^{1,2}$, S.~W.~Duchesne$^{3,5}$,
A.~Bonafede$^{1,2}$, L.~Rudnick$^{10}$, M.~Hoeft$^{11}$, B.~Quici$^{5}$, D.~Eckert$^{12}$, 
\newauthor 
M.~Brienza$^{1,2}$, C.~Tasse$^{13,14}$, E.~Carretti$^{2}$, J.~D.~Collier$^{15,16,3}$,
J.~M.~Diego$^{17}$, L.~Di~Mascolo$^{18,19,20}$,
\newauthor
A.~M.~Hopkins$^{21}$,  M.~Johnston-Hollitt$^{22}$, R.~R.~Keel$^{5}$, B.~S.~Koribalski$^{8,16}$, T.~H.~Reiprich$^{23}$
\\
(Affiliations can be found after references)
}

\date{Accepted 2022~June~20. Received 2022~June~13; in original form 2022~May~4
}

\pubyear{2022}

\begin{document}
\label{firstpage}
\pagerange{\pageref{firstpage}--\pageref{lastpage}}
\maketitle

\begin{abstract}
Abell~3266 is a massive and complex merging galaxy cluster that exhibits significant substructure. We present new, highly sensitive radio continuum observations of Abell~3266 performed with the Australian Square Kilometre Array Pathfinder (0.8$-$1.1 GHz) and the Australia Telescope Compact Array (1.1$-$3.1 GHz). These deep observations provide new insights into recently-reported diffuse non-thermal phenomena associated with the intracluster medium, including a ‘wrong-way’ relic, a fossil plasma source, and an as-yet unclassified central diffuse ridge, which we reveal comprises the brightest part of a large-scale radio halo detected here for the first time. The ‘wrong-way’ relic is highly atypical of its kind: it exhibits many classical signatures of a shock-related radio relic, while at the same time exhibiting strong spectral steepening. While radio relics are generally consistent with a quasi-stationary shock scenario, the ‘wrong-way’ relic is not. We study the spectral properties of the fossil plasma source; it exhibits an ultra-steep and highly curved radio spectrum, indicating an extremely aged electron population. The larger-scale radio halo fills much of the cluster centre, and presents a strong connection between the thermal and non-thermal components of the intracluster medium, along with evidence of substructure. Whether the central diffuse ridge is simply a brighter component of the halo, or a mini-halo, remains an open question. Finally, we study the morphological and spectral properties of the multiple complex radio galaxies in this cluster in unprecedented detail, tracing their evolutionary history.
\end{abstract}


\section{Introduction}
Residing at the intersections of the large-scale structure of the Universe's Cosmic Web, galaxy clusters are some of the largest gravitationally-bound objects in the Universe. Around $\sim80$~per~cent of the mass of a galaxy cluster is comprised of dark matter, with the remaining $\sim20$~per~cent being baryonic matter. The dominant contribution to this baryonic matter content of a cluster is the intracluster medium (ICM), a hot ($T \sim10^7 - 10^8$~K), tenuous ($n_{\rm{e}} \sim 10^{-3}$~cm$^{-3}$) plasma which emits via the bremsstrahlung mechanism at X-ray wavelengths and scatters photons of the Cosmic Microwave Background (CMB) via the Sunyaev-Zeldovich effect.

In the hierarchical model of structure formation, galaxy clusters grow through a variety of processes, including accretion of material from the external medium, absorption of individual galaxies and groups, and major merger events. Major mergers are some of the most dramatic events since the Big Bang, depositing tremendous quantities of energy \citep[$\sim10^{64}$~erg;][]{Ferrari2008} into the ICM in the form of shocks and turbulence.

Spectacular sources of diffuse radio emission have been detected in many merging clusters \citep[for a recent observational review, see][]{vanWeeren2019}. These sources of diffuse emission fall broadly into three categories: radio relics, radio haloes, and mini-haloes. However, new and highly sensitive low-frequency radio observations have revealed that the picture is much more complex, and other types of diffuse source (many of which do not fit our taxonomy) have been detected in recent years \citep[e.g.][]{Govoni2019,Botteon2020c,Botteon2020b,Ignesti2020b,Hodgson2021,Knowles2022,Venturi2022}. The extremely large spatial scale of these sources, combined with the short lifetime of the relativistic electrons generating the synchrotron emission ($\tau\lesssim10^8$~yr) implies in-situ acceleration by physical processes in the ICM.

Radio relics are extended synchrotron sources (typically megaparsec-scale) that are usually located toward the cluster outskirts. When imaged at sufficiently high resolution, relics frequently show significant filamentary substructure \citep[for example][]{Owen2014,vanWeeren2017b,DiGennaro2018,Rajpurohit2018,Rajpurohit2020a,Botteon2020b,deGasperin2022}. Relics also frequently exhibit a high linear polarisation fraction, with magnetic field vectors aligned with the long axis of the relic, presumably tracing shock compression of magnetic fields \citep[e.g.][]{Bonafede2012,Riseley2015,Loi2020,Rajpurohit2021c,deGasperin2022}.

Relics are generally believed to be associated with shocks induced by merger events, and indeed a growing number of clusters show associations between relics and shocks detected at X-ray wavelengths \citep[][]{Finoguenov2010,Shimwell2015,Botteon2016b}. It is commonly accepted that the mechanism responsible for deposition of shock energy into the ICM is some form of diffusive shock acceleration \citep[DSA;][]{Ensslin1998,Roettiger1999}. However, one of the open questions is the origin of the cosmic ray electron (CRe) population that gives rise to the synchrotron emission. 

Specifically, the most active area of debate is whether the CRe population originates from the thermal pool \citep[][]{Hoeft2007} or from a pre-existing population of mildly-relativistic electrons \citep[][]{Kang2011}, which is frequently referred to as the `re-acceleration' scenario. Indeed, a growing number of radio relics have been found which appear to show connections with cluster-member radio galaxies, providing support for the re-acceleration scenario \citep[][]{vanWeeren2017,Stuardi2019,HyeongHan2020,Jones2021}.

While `standard' DSA from the thermal pool can successfully recreate many observed properties of radio relics, some major difficulties remain. In the case of weak shocks, DSA requires an unphysically large acceleration efficiency to reproduce the most luminous radio relics \citep{Botteon2020a}. Additionally, the Mach numbers derived from X-ray observations are frequently found to be lower than those derived from radio observations. However, recent simulations of cluster mergers can go some way to reconciling this discrepancy between radio-derived and X-ray-derived Mach numbers, depending on the system geometry \citep{Hong2015,Dominguez-Fernandez2021,Wittor2021}. Furthermore, the theoretical framework of the multi-shock scenario --- whereby an existing electron population is re-accelerated by repeated shocks --- is now being explored \citep{Kang2021,Inchingolo2022}. The CRe population generated by this scenario shows significantly higher luminosity than that of a CRe population that has been shocked only a single time; in addition, the multi-shock CRe population shows a spectral energy distribution (SED) that clearly departs from a single power-law behaviour.

Radio haloes are centrally-located, largely amorphous, megaparsec-scale synchrotron sources that usually exhibit negligible polarisation fraction. The synchrotron emission from haloes generally follows the distribution of the thermal ICM (as traced by X-ray emission), implying a direct connection between thermal and non-thermal components in the ICM \citep[see][for example]{Govoni2001a,Govoni2001b,Feretti2001,Giovannini2009,Boschin2012,Shimwell2014,Cantwell2016,Rajpurohit2021b,Rajpurohit2021c,Bruno2021,deGasperin2022}.

The more commonly accepted scenario for generation of radio haloes involves the acceleration of CRe through turbulence injected into the ICM following merger events \citep{Brunetti2001,Petrosian2001}. The turbulent (re-)acceleration scenario naturally links the presence of radio haloes with ongoing cluster mergers, and successfully explains many of the observed halo properties, such as the existence of ultra-steep spectrum radio haloes \citep[e.g.][]{Brunetti2008}.

One alternative to the turbulent (re-)acceleration scenario is the secondary (or hadronic) scenario, whereby CRe are injected throughout the ICM via inelastic collisions between cosmic ray protons (CRp) and thermal protons of the ICM \citep{Dennison1980,Blasi1999}. However, purely-hadronic scenarios cannot work for radio haloes: even the recent claim of a gamma-ray detection toward the Coma cluster \citep{Xi2018}, if attributed \emph{entirely} to hadronic collisions in the ICM, only accounts for up to 20\% of the observed radio halo luminosity \citep{Adam2021}.

Finally, mini-haloes remain the most poorly-understood class, as only some 32 are currently known, compared to around 100 radio relics and a similar number of radio haloes \citep{vanWeeren2019}. Mini-haloes are moderately-extended (typically around $0.1-0.3$~Mpc) objects that are centrally-located in relaxed clusters which also host a powerful radio source associated with the brightest cluster galaxy (BCG).

Although it remains an active area of theoretical discussion, two principal formation mechanisms exist for mini-haloes: either turbulent (re-)acceleration  \citep{Gitti2002,ZuHone2013,ZuHone2015} or hadronic processes \citep{Pfrommer2004}. Under the former scenario, mini-haloes are generated by synchrotron radiation from CRe (possibly injected by the BCG radio counterpart as well as other cluster-member radio galaxies) which have been (re-)accelerated to relativistic energies by minor/off-axis merger events, core sloshing within the cool core and/or gas dynamics driven by cold fronts \citep[see for example][]{Gitti2002,ZuHone2013,Brunetti2014}. However, any minor merger event would have to generate large-scale bulk motions of gas while leaving the cool core intact \citep{ZuHone2015}.

The hadronic scenario is much the same as that for radio haloes: that mini-haloes are generated by relativistic electrons formed through collisions between CRp and thermal protons of the ICM. While tensions with limits from \emph{Fermi} also exist for mini-haloes, recent work by \cite{Ignesti2020} showed that the limits do not completely exclude the hadronic scenario. Recent multi-frequency observations of mini-haloes have shown that our understanding of their underlying physics is far from complete, with multi-component mini-haloes being found in increasing numbers \citep{Savini2018,Savini2019,Raja2020,Biava2021,Riseley2022}. In many of these cases, both the turbulent and hadronic mechanisms can explain some of the observational evidence, but neither scenario is able to explain all physical properties, and thus further work is required to understand the detailed and complex physical processes at work.

As well as these classes of diffuse synchrotron emission that originate with physical processes in the ICM, many clusters also host embedded resolved radio galaxies that exhibit a range of morphologies. We will collectively refer to these sources as `bent-tailed radio galaxies' as further sub-categorisation --- such as the canonical wide-angle-tail (WAT), narrow-angle-tail (NAT), and head-tail (HT) classifications --- is highly subjective and strongly affected by a number of observational systematics (such as projection effects). 

The morphological and spectropolarimetric properties of bent-tailed radio galaxies provide powerful probes of the cluster weather, as well as thermal and non-thermal components in the ICM \citep[e.g.][]{Burns1998,Pratley2013,JohnstonHollitt2015b,Lal2020,GendronMarsolais2020,GendronMarsolais2021,Brienza2021}. For example, turbulence, cluster winds, bulk motions and/or the relative motion of these galaxies can all result in a variety of different observed morphologies.

Remnant radio galaxies (also known as `dying' or `fossil' radio sources) represent the end phase of radio galaxy evolution. These sources comprise aged plasma left behind by an active galaxy after its central engine has switched off. Following standard synchrotron ageing processes, these sources fade more rapidly toward higher frequencies, becoming detectable only at increasingly long wavelengths before fading altogether \citep[e.g.][]{Murgia2011,Harwood2013,Harwood2015,Brienza2016,Godfrey2017,Turner2018a,Turner2018b,Quici2022}.

With the advent of the next-generation long wavelength interferometers such as the LOw-Frequency ARray \citep[LOFAR;][]{vanHaarlem2013} and the Murchison Widefield Array \citep[MWA;][]{Tingay2013,Wayth2018}, our understanding of fossil plasma sources is increasing. However, despite them being studied in increasing numbers \citep[for example][]{Brienza2017,Mahatma2018,Duchesne2019,Randriamanakoto2020,Hodgson2021,Jurlin2021,Brienza2021,Brienza2022,Knowles2022}, they remain a relatively rare and short-lived phase of radio galaxy evolution, comprising some $\sim5-15$ per cent of the radio galaxy population.

\subsection{Abell 3266}
Situated at (J2000 Right Ascension, Declination) = (04h31m24.1s, $-$61d26m38s), Abell~3266 (also known as Sersic 40$-$6) is a nearby cluster of galaxies at redshift $z=0.0594$ \citep{Quintana1996}. With a total mass of $(6.64^{+0.11}_{-0.12})\times10^{14}$~M$_{\odot}$ \citep{Planck2016b}, it is the most massive member of the Horologium-Reticulum supercluster \citep{Fleenor2005}.

Abell~3266 has been extensively observed at X-ray wavelengths \citep{Markevitch1998,Mohr1999,DeGrandi1999,Henriksen2000,Sauvageot2005,Finoguenov2006,Sanders2022}. All observational results show that Abell~3266 is undergoing a complex merger event, presenting an elongated and asymmetric X-ray surface brightness distribution, as well as an asymmetric temperature profile. Most recently, \cite{Sanders2022} reported a detailed X-ray analysis of Abell~3266 using new data from the eROSITA telescope \citep{Predehl2021} on board the \emph{Spectrum R\"{o}ntgen Gamma} (\emph{SRG}) satellite. eROSITA revealed much substructure within the cluster, including two weak (Mach number $\mathcal{M} \sim 1.5-1.7$) shock fronts toward the cluster periphery at a distance of around $1.1$~Mpc from the cluster centre.

Detailed optical observations presented by \cite{Dehghan2017} have revealed that the cluster contains some $\sim800$ member galaxies, distributed between two sub-clusters and at least six sub-groups (which lie predominantly to the North and NW of the cluster centre). Overall, Abell~3266 exhibits a significant velocity dispersion of ${\rm{v}}_{\rm{d}} \sim 1400$~km~s$^{-1}$ \citep{Dehghan2017}, although those authors note that this is likely due to the relative motions of the various sub-clusters and sub-groups comprising this complex system. The BCG exhibits a dumbbell morphology and shows strong tidal distortion, likely arising as a result of the merger between the two main sub-clusters \citep{Henriksen2000,Henriksen2002}.

Despite the extensive X-ray and optical coverage, there is a comparative lack of detailed radio observations. Most historic studies have been performed with either narrow-band instruments \citep[][]{1999PhDTReid,1999Murphy,Murphy2002} or low-resolution radio telescopes \citep[][]{RiseleyThesis,Bernardi2016}. Analysis of narrow-band observations performed with the Australia Telescope Compact Array \citep[ATCA;][]{FraterBrooks1992} reveals the cluster is host to several spectacular tailed radio galaxies as well as a number of compact and diffuse sources \citep{1999Murphy,1999PhDTReid,DehghanThesis}. One of the extended sources, situated to the NW of the cluster centre, shows an elongated morphology and possesses a steep spectrum between 1.4\,GHz and 843\,MHz. When viewed in concert with a potential optically-selected host galaxy, the interpretation was that this elongated radio source is perhaps a remnant radio galaxy \citep{1999Murphy} or a radio relic \citep{DehghanThesis}.

Abell~3266 was observed with MeerKAT as one of the X-ray selected clusters in the MeerKAT Galaxy Cluster Legacy Survey \citep[GCLS;][]{Knowles2022}, under the identifier J0431.4$-$6126. These observations have been used to study the resolved nature of the aforementioned complex radio galaxy, detecting a variety of features not seen before in this source. For highlights, see \cite{Knowles2022}; for a detailed discussion and interpretation of the features seen in this source, see \cite{Rudnick2021}. \cite{Knowles2022} also report the first clear detection of a radio relic in Abell~3266, although \cite{1999Murphy} reported a tentative identification at around $2.5\sigma$ significance.

\cite{Duchesne2022} present the first low-frequency observations of Abell~3266 using the Phase II MWA \citep{Wayth2018}, covering the frequency range 88$-$216~MHz. These observations confirm the detection of the radio relic, as well as revealing two ultra-steep-spectrum fossil plasma sources in the ICM of Abell~3266. However, while Abell~3266 bears many of the typical characteristics of clusters that host a radio halo, the deep MWA observations presented by \cite{Duchesne2022} do not detect a halo. By injecting simulated haloes with a variety of spatial scales and luminosities, \citeauthor{Duchesne2022} demonstrate that any halo that is present but undetected would lie up to a factor five below established scaling relations between cluster mass and radio halo power \citep[e.g.][]{Cassano2013,Duchesne2021b}.

\begin{figure*}
\begin{center}
\includegraphics[width=1.0\textwidth]{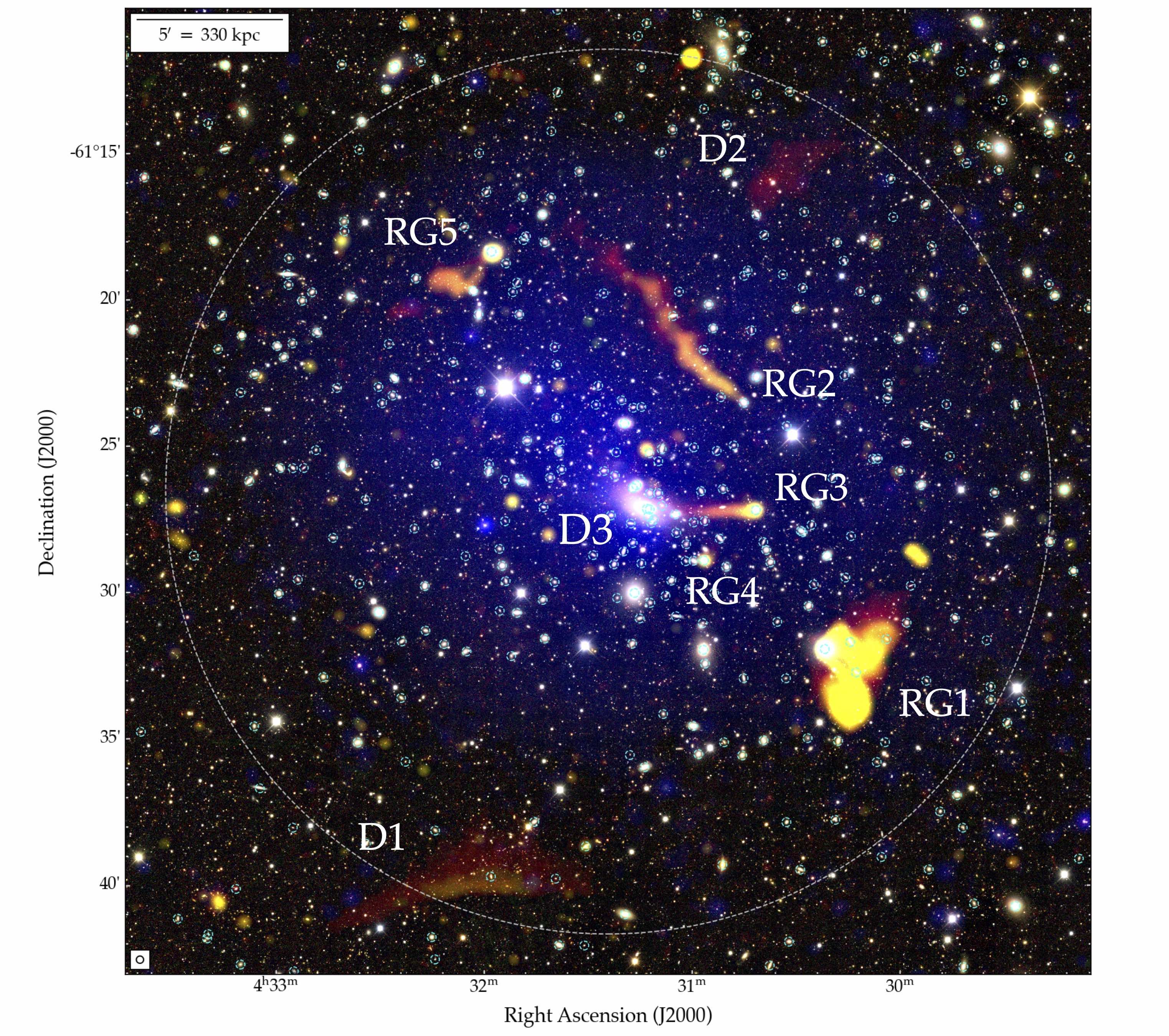}
\cprotect\caption{Colour-composite image of Abell 3266, overlaying radio colour on optical RGB. Optical RGB image comprises $i$-, $r$- and $g$-bands from Data Release 2 of the Dark Energy Survey (DES). We overlay the radio surface brightness measured by ASKAP (943~MHz; red channel) and the ATCA (2.1~GHz; green channel), both at 15~arcsec resolution. Also overlaid is the [0.5$-$2~keV] X-ray surface brightness measured by \xmm{} (blue channel). Dashed cyan circles indicate cluster-member radio galaxies identified by \protect\cite{Dehghan2017}. The dashed silver circle indicates a 1~Mpc radius from the cluster centre, given our cosmology. Labels identify sources of interest to be discussed later, with `D' indicating that a source's nature is `diffuse', and `RG' indicating that it is an active radio galaxy.}
\label{fig:A3266_RGB}
\end{center}
\end{figure*}

This paper presents analysis of new, deep, broad-band radio data from the ATCA and the Australian Square Kilometre Array Pathfinder \citep[ASKAP;][]{Johnston2007,DeBoer2009,Hotan2021}, in conjunction with ancillary X-ray data from \xmm{}. These observations were performed independently of the MeerKAT GCLS data release; the results are very complementary and allow us to gain new and detailed insights into this rich, complex cluster. We present our showcase multi-wavelength image of Abell~3266 in Figure~\ref{fig:A3266_RGB}.

The remainder of this paper is divided as follows: we discuss the observations and data reduction in \S\ref{sec:observations}, we present our results in \S\ref{sec:results}. In \S\ref{sec:diffuse} we discuss our new results on diffuse radio sources in Abell~3266, and in \S\ref{sec:tailed_galaxies} we discuss selected tailed radio galaxies. We draw our conclusions in \S\ref{sec:conclusions}. Throughout this work we take the redshift of Abell 3266 to be $z=0.0594$ \citep{Quintana1996}. We assume a $\Lambda$CDM cosmology of H$_0 = 73 \, \rm{km} \, \rm{s}^{-1} \, \rm{Mpc}^{-1}$, $\Omega_{\rm{m}} = 0.27$, $\Omega_{\Lambda} = 0.73$. With this cosmology, at the redshift of Abell 3266, an angular distance of 1\,arcsec corresponds to a physical distance of 0.906\,kpc. We adopt the convention that flux density $S$ is proportional to observing frequency $\nu$ as $S \propto \nu^{\alpha}$ where $\alpha$ is the spectral index, and all uncertainties are quoted at the $1\sigma$ level.

\section{Observations \& data reduction}\label{sec:observations}

\subsection{Australia Telescope Compact Array}
We observed Abell 3266 with the ATCA between 2020 October and 2021 October, for a total of approximately 66~hours (project ID C3389, P.I. Riseley). Observations were performed using the Compact Array Broad-band Backend \citep[CABB;][]{Wilson2011} in the 16\,cm band, which covers the frequency range $1.1-3.1$~GHz at a spectral resolution of 1\,MHz.

Four complementary array configurations were used, with baselines between 30\,m and 6\,km in length, providing sensitivity to a broad range of angular scales and a nominal resolution of around 6\,arcsec. We covered Abell 3266 using a mosaic of four pointings, spaced at $50$\,per\,cent of the primary beam (PB) full-width-at-half-maximum (FWHM), in order to ensure that the entire cluster volume could be probed in full polarisation while minimising the effects of off-axis instrumental leakage. This paper presents our continuum results; polarisation data will be presented in a future paper. Our observations are summarised in Table~\ref{tab:obs_summary}.

\begin{table}
\footnotesize
\centering
\caption{ATCA observations of Abell 3266. \label{tab:obs_summary}}
\begin{tabular}{lccccp{3mm}p{6.5cm}}
\hline
Date 	     &  Start time & Hours    & Array          \\
      	     &             & Observed & Configuration  \\
 	   	     & (AEST)      &          &                \\
\hline 
2020 Oct. 24 & 19:30       & 14       & 6B    \\
2020 Oct. 25 & 19:30       & 14       & 6B    \\
2020 Dec. 18 & 17:00       & 14       & 1.5A \\
2020 Dec. 20 & 21:30       & 6        & 1.5A \\
2021 Jan. 26 & 14:30       & 12       & EW352 \\
2021 Oct. 10 & 17:30       & 5.5      & H168 \\
\hline
\end{tabular}
\end{table}

Our flux density scale was set using a 10$-$15 minute scan of PKS~B1934$-$638 on each day. We applied standard calibration and radio frequency interference (RFI) flagging techniques using the \textsc{miriad} software package \citep{Sault1995}. A total of around $40$\,per\,cent of the data were flagged due to the presence of strong RFI, which dominates specific channel ranges and occurs primarily at the low-frequency end of the band.

All imaging was carried out using the \textsc{ddfacet} software package \citep{Tasse2018}. We employed \texttt{robust} $=0$ weighting \citep{Briggs1995} to strike a balance between sensitivity and resolution. As Abell 3266 hosts a variety of extended radio sources, we used the sub-space deconvolution algorithm \citep[\textsc{ssd};][]{Tasse2018} in order to more correctly model this diffuse radio emission.

Self-calibration was performed using the \textsc{killms} software package \citep{Tasse2014,Smirnov2015}. Initial self-calibration was carried out in direction-independent (DI) mode; four rounds of phase-only self-calibration were performed, followed by three rounds of amplitude \& phase self-calibration. All four pointings were jointly deconvolved using new functionality incorporated into the \verb|dev| version of \textsc{ddfacet} for this project, including implementation of the ATCA 16~cm band beam model\footnote{See \url{https://www.narrabri.atnf.csiro.au/people/ste616/beamshapes/beamshape_16cm.html} for details of the ATCA 16~cm band beam model.}.

We used the quality-based weighting scheme developed by \cite{Bonnassieux2018} to weight our calibration solutions throughout our self-calibration process. This weighting scheme uses the gain statistics (which are used within the calibration solver to find optimal gain solutions) to estimate the relative quality of each gain solution, and weighs the corrected visibilities appropriately. After each solve, \textsc{killms} computes these weights by default, as minimal computational overhead is required. Typically, usage of this weighting scheme results in faster convergence of the self-calibration process.

After seven rounds of DI self-calibration, our solutions had largely converged, and we then inspected our mosaic for residual direction-dependent (DD) errors. While the quality of our jointly-deconvolved mosaic was generally high, some of the brighter sources in Abell~3266 showed residual DD errors, limiting our overall dynamic range. We tessellated the sky into twelve clusters, and performed several rounds of DD self-calibration and imaging, solving for gains in both amplitude and phase and refining our solutions with each iteration. Once our DD self-calibration had suitably converged, we proceeded to generate our final science images. The native resolution of our \texttt{robust} $=0$ ATCA mosaic is around 4.5~arcsec; to match the resolution of our ASKAP mosaics, we generated our science images at 9~arcsec resolution and 15~arcsec resolution by varying the \texttt{robust} parameter appropriately (\texttt{robust} $=0$ for our 9~arcsec image, \texttt{robust} $=+0.7$ for our 15~arcsec image) before convolving with an elliptical Gaussian to achieve the desired resolution. We adopt a representative 5 per cent calibration uncertainty in our ATCA mosaic of Abell~3266. The right panel of Figure~\ref{fig:A3266_fullres} shows our 6~arcsec ATCA image of Abell~3266; Figure~A1 of the supplementary material illustrates the dramatic improvement in image fidelity enabled by applying these next-generation algorithms to ATCA data processing.

\subsection{Australian Square Kilometre Array Pathfinder}\label{sec:askap}
ASKAP comprises 36 twelve-metre dishes located in the Murchison Radioastronomy Observatory (MRO) in Western Australia, observing between 700\,MHz and 1.8\,GHz, with an instantaneous bandwidth of up to 288\,MHz. Due to the unique phased-array feed \citep[PAF;][]{Hotan2014,McConnell2016} technology, ASKAP is capable of simultaneously forming up to 36 independent beams, covering some $\sim30$\,deg$^2$, making ASKAP a powerful survey telescope.

With this large field of view (FOV), operating at low frequencies, and with baselines in the range 22\,m and 6\,km (providing sensitivity to radio emission on a wide range of angular scales), ASKAP is ideally suited to observations of low surface brightness diffuse radio sources typically found in many clusters of galaxies \citep{HyeongHan2020,Wilber2020,Bruggen2021,DiMascolo2021,Duchesne2021a,Duchesne2021b,Venturi2022}.

Abell 3266 was observed with ASKAP during Early Science operations for the Evolutionary Map of the Universe \citep[EMU;][]{Norris2011} survey, under SBID~10636. Observations were performed on 2019~Nov.~24, for a total on-source time of approximately seven hours. These observations were performed with 34 antennas, as two (ak09, ak32) did not participate in the observing run. The field was observed in Band 1 at a central frequency of 943\,MHz, with a bandwidth of 288\,MHz. Our target was observed using the `closepack-36' beam configuration, covering a total FOV of around 30\,deg$^2$.

PKS~B1934$-$638 is routinely used by ASKAP for bandpass calibration and absolute flux density scale calibration, and is observed daily on a per-beam basis. Initial data processing was performed using the \textsc{askapsoft} pipeline \citep{Guzman2019}, which incorporates RFI flagging, bandpass calibration, averaging and multiple rounds of DI amplitude and phase self-calibration\footnote{\url{https://www.atnf.csiro.au/computing/software/askapsoft/sdp/docs/current/pipelines/introduction.html}}. Each beam is processed independently before being co-added in the image plane. Following pipeline processing, the data were averaged to 1\,MHz spectral resolution. Aside from the two missing antennas, around 15\,per\,cent of the data were flagged by the \textsc{askapsoft} pipeline.

The mosaic has a common resolution of $15.5~{\rm arcsec}~\times13.7$~arcsec, with an off-source rms noise of $33\,\upmu$Jy beam$^{-1}$ in the vicinity of Abell 3266. In general, the quality of the full ASKAP mosaic covering the Abell 3266 field was very high. However, residual DD errors remain in the mosaic -- particularly evident in regions around relatively bright sources $(S \gtrsim 1~{\rm{Jy}})$. As such, we opted to perform postprocessing using \textsc{ddfacet} and \textsc{killms}.

Abell 3266 was principally covered by two beams (Beam 15 and Beam 20) from the ASKAP mosaic. As such, we retrieved the calibrated visibilities for these beams via the CSIRO\footnote{Commonwealth Scientific \& Industrial Research Organisation} ASKAP Science Data Archive \citep[CASDA\footnote{\url{https://research.csiro.au/casda/}};][]{Chapman2017,Huynh2020}. These beams were processed independently, and mosaicked in the image plane after performing further self-calibration.

For each beam, we imaged an area of $3\times3$ square degrees, in order to include particularly bright sources in the wide field. Imaging was performed in a similar manner to our ATCA data, using \textsc{ddf} with the \textsc{ssd} algorithm. We performed a single round of DI self-calibration, before switching to the DD regime. The sky was tessellated into nine clusters, and we performed two rounds of amplitude-and-phase self-calibration and imaging on each beam.

To correct for PB attenuation, we followed the same principles applied by \cite{Wilber2020}. We assume a Gaussian PB with a FWHM of the form ${\rm{FWHM}} = 1.09 \lambda/D$, where $D=12$~metres is the antenna diameter, and $\lambda=0.3179$~metres is the wavelength at our reference frequency (943~MHz). We linearly mosaicked our final DD-calibrated images of our two selected ASKAP beams, weighted according to the PB response, and applied a factor $\times2$ to account for the difference between the Stokes $I$ convention adopted by ASKAP and other imaging software. We also applied a bootstrap factor of 1.075 to our ASKAP data, as suggested by the CASDA validation report; as such, we adopt a conservative 10 per cent calibration uncertainty in our ASKAP maps of Abell~3266.

We show our final ASKAP image of Abell~3266 at 15~arcsec resolution in the left panel of Figure~\ref{fig:A3266_fullres}. We also demonstrate the significant improvement in image fidelity that is enabled by applying \textsc{killms} and \textsc{ddfacet} to ASKAP data in Figure~A2 of the supplementary material. We note that the majority of the large scale `diffuse flux' in the left panel of Figure~A2 is a manifestation of calibration errors associated with the DI processing of these data. The significant suppression in these artefacts brought about by DD calibration and our use of \textsc{ssd}-clean is evident in the right panel of Figure~A2.

\begin{table}
\footnotesize
\centering
\caption{\xmm{} observations of Abell 3266. \label{tab:xmm_summary}}
\begin{tabular}{lcccc}
\hline
ObsID & Target & \multicolumn{3}{c}{Clean exposure time (ks)} \\
& & MOS1 & MOS2 & pn \\
\hline 
0105260701 & A3266\_f1 & 19.0 & 19.3 & 15.5 \\ 
0105260801 & A3266\_f2 & 19.7 & 19.7 & 15.5 \\ 
0105260901 & A3266\_f3 & 23.4 & 23.0 & 17.9 \\ 
0105261001 & A3266\_f4 & 2.8 & 2.8 & 0.5 \\ 
0105261101 & A3266\_f5 & 11.6 & 12.1 & 7.1 \\ 
0105262001 & A3266\_f6 & 6.2 & 5.7 & 2.5 \\ 
0105262101 & A3266\_f5 & 5.6 & 6.2 & 2.5 \\
0105262201 & A3266\_f4 & 2.9 & 2.9 & 3.2 \\
0105262501 & A3266\_f6 & 6.2 & 6.9 & 3.2 \\
0823650301 & A3266NW & 22.3 & 22.9 & 10.4 \\
0823650401 & A3266NE & 24.3 & 26.0 & 16.1 \\
0823650601 & A3266SW & 24.2 & 25.4 & 20.2 \\
0823650701 & A3266NW & 31.2 & 32.8 & 20.3 \\
\hline
\end{tabular}
\end{table}

\subsection{\xmm{}}
We retrieved from the \xmm{} Science Archive\footnote{\url{http://nxsa.esac.esa.int/nxsa-web}} the 13 observations on Abell 3266 that are listed in Tab.~\ref{tab:xmm_summary}. These data were collected in three different epochs: ObsIDs 0105260701$\dots$0105262201 in 2000, ObsID 0105262501 in 2003, and ObsIDs 0823650301$\dots$0823650701 in 2018. While the earliest observations cover the central region of the cluster and were originally presented by \citet{Sauvageot2005} and \citet{Finoguenov2006}, the latest observations cover the virial radius region and were carried out in the context of the \xmm{} Cluster Outskirts Project \citep[X-COP;][]{Eckert2007}. We processed the data using the \xmm{} Scientific Analysis System (SAS v16.1.0) and the Extended Source Analysis Software (ESAS) data reduction scheme \citep{Snowden2008} following the working flow described by \citet{Ghirardini2009}. The final clean exposure for each ObsID and for each detector of the European Photon Imaging Camera (EPIC) is summarized in Tab.~\ref{tab:xmm_summary}. We then combined the count images and corresponding exposure images of each dataset to produce a mosaic exposure-corrected image in the 0.5$-$2.0 keV band that was used to compare the distribution of the synchrotron diffuse cluster sources with respect to the thermal ICM.

\section{Results}\label{sec:results}
Figure~\ref{fig:A3266_RGB} presents our `showcase' colour-composite image of Abell 3266, with radio colour overlaid on an optical RGB image constructed from Data Release 2 (DR2) of the Dark Energy Survey \citep[DES;][]{DES-DR2}. The optical RGB image uses DES $i$-, $r$- and $g$-bands in the respective channels. The radio surface brightnesses measured from ASKAP and ATCA maps (both at 15~arcsec resolution) are inserted in the red and green channels, respectively. Finally, we show the X-ray surface brightness measured by \xmm{}, smoothed using a Gaussian filter with a 7~pixel FWHM, in the blue channel. In this image, typical synchrotron emission ($\alpha \sim -0.8$) exhibits a yellow hue, with steep-spectrum emission appearing redder in colour. Conversely, flat-spectrum sources appear green in colour.

\begin{figure*}
\begin{center}
\includegraphics[width=1.0\textwidth]{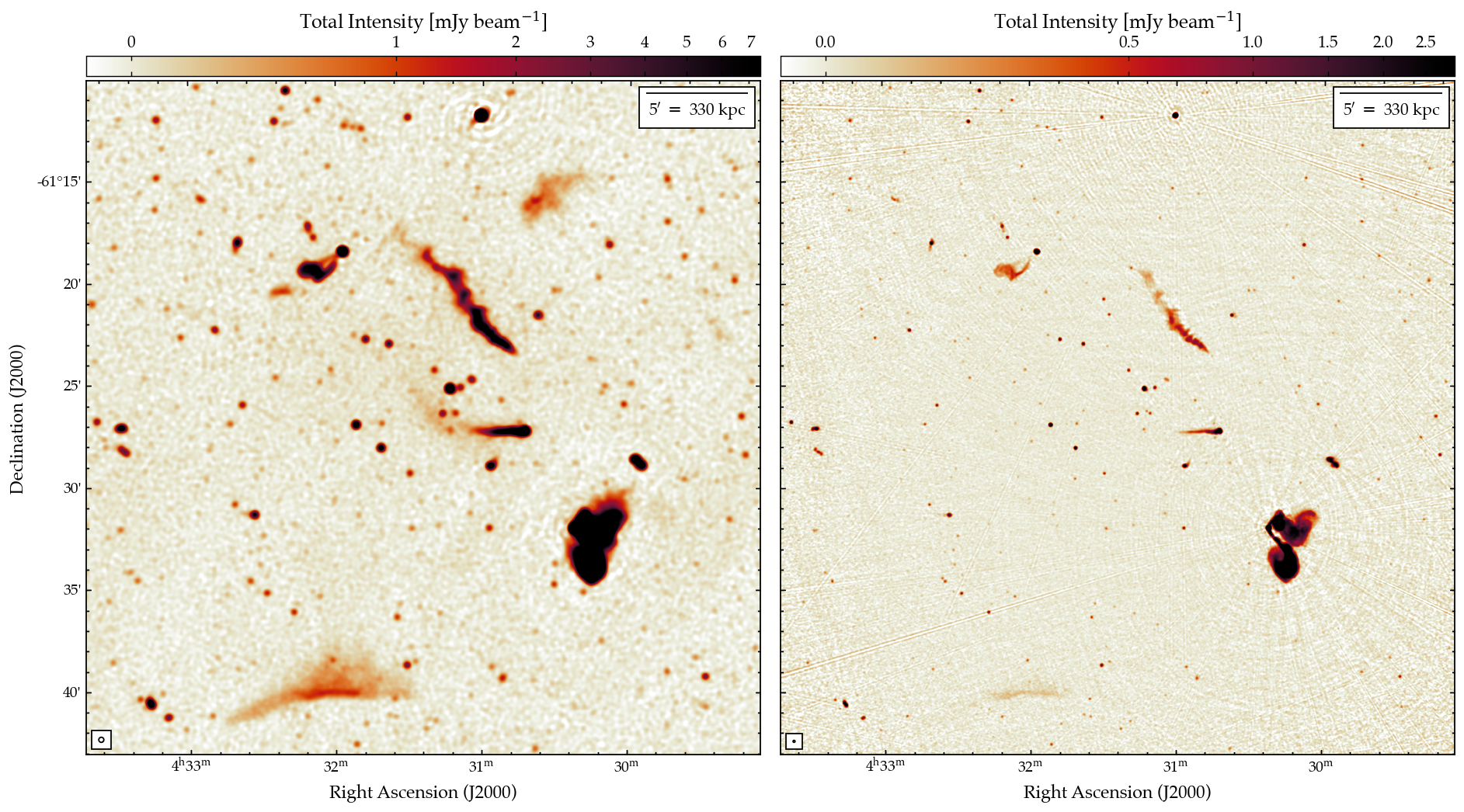}
\cprotect\caption{Example radio images of Abell 3266, as observed with ASKAP (\emph{left}; 943~MHz) and the ATCA (\emph{right}; 2.1~GHz). Colour maps range from $-3\sigma$ to $200\sigma$ on an arcsinh stretch, to emphasise faint, diffuse emission. The typical off-source noise is $36.7~\upmu$Jy beam$^{-1}$ in our ASKAP image (15~arcsec resolution), and $14.5~\upmu$Jy beam$^{-1}$ in our ATCA image ($6$~arcsec resolution).}
\label{fig:A3266_fullres}
\end{center}
\end{figure*}

We also present example radio maps in Figure~\ref{fig:A3266_fullres}. The ASKAP image shown in Figure~\ref{fig:A3266_fullres} has an rms noise of $36.7~\upmu$Jy beam$^{-1}$ at a resolution of 15~arcsec and a reference frequency of 943~MHz. The ATCA image displayed in Figure~\ref{fig:A3266_fullres} has an rms noise of $14.5~\upmu$Jy beam$^{-1}$ at a resolution of 6~arcsec and a reference frequency of 2.1~GHz. 

Our new observations reveal a plethora of spectacular tailed radio galaxies along with three diffuse sources, the nature of which is not immediately obvious. All analysis detailed later in this paper is performed on matched-resolution images at 9~arcsec and 15~arcsec. We opted to study the radio properties of the cluster at two different resolutions in order to better recover the spatially-resolved properties of the complex cluster-member radio galaxies (by using our 9~arcsec resolution images) and improve our signal-to-noise (S/N) when studying the diffuse, steep-spectrum non-thermal sources (using our 15~arcsec resolution images). All image properties (such as RMS noise and resolution) are summarised in Table~\ref{tab:img_summary}. 

\begin{table}
\footnotesize
\centering
\caption{Summary of image properties for our maps of Abell~3266. `LAS' denotes the theoretical largest recoverable angular scale of each image. \label{tab:img_summary}}
\begin{tabular}{lp{1cm}rrr}
\hline
Image         &  Imaging \texttt{robust} & RMS noise              & Resolution & LAS      \\
              &                          & [$\upmu$Jy beam$^{-1}$]  & [arcsec]   & [arcmin] \\
\hline 
\multirow{2}{*}{ASKAP} & $-1.5$ & 36.7                     &  15        &  \multirow{2}{*}{48}  \\
      & $-2.0$ & 90.9                     &   9        &           \\
\multirow{2}{*}{ATCA}       & $+0.7$ & 13.9                     &  15        &  \multirow{2}{*}{16}  \\
       & $+0.0$ & 13.7                     &   9        &           \\
\hline
\end{tabular}
\end{table}

\section{Diffuse radio sources}\label{sec:diffuse}
Throughout this analysis, we draw predominantly on the ASKAP and ATCA images presented in this work. Where appropriate, we incorporate flux density measurements from the MWA-2 data presented by \cite{Duchesne2022}. We also incorporate data products from MeerKAT images of Abell~3266 from the recent first data release from the MeerKAT GCLS \citep{Knowles2022}. These data products have not been corrected for DD effects, and as such the image fidelity and dynamic range varies with frequency; this is principally visible at the lower end of the MeerKAT band, where a bright off-axis radio source far to the south-west of Abell~3266 causes artefacts which underlay much of the cluster\footnote{This source was also problematic for the DI-calibrated ASKAP mosaic; after DD-calibration, the impact of this far-field source on our mosaic is negligible.}. However, at frequencies above around 1~GHz, these effects become increasingly suppressed due to the narrowing PB FWHM. As such, we extracted images at frequencies of 1043, 1092, 1312, 1381, 1448, 1520, and 1656~MHz from the MeerKAT GCLS 15~arcsec resolution cube, and incorporated these into our analysis. We assume a representative 10 per cent calibration uncertainty for all MeerKAT measurements incorporated in our analysis.

\subsection{Source D1: a `wrong-way' relic}
\begin{figure}
\begin{center}
\includegraphics[width=\linewidth]{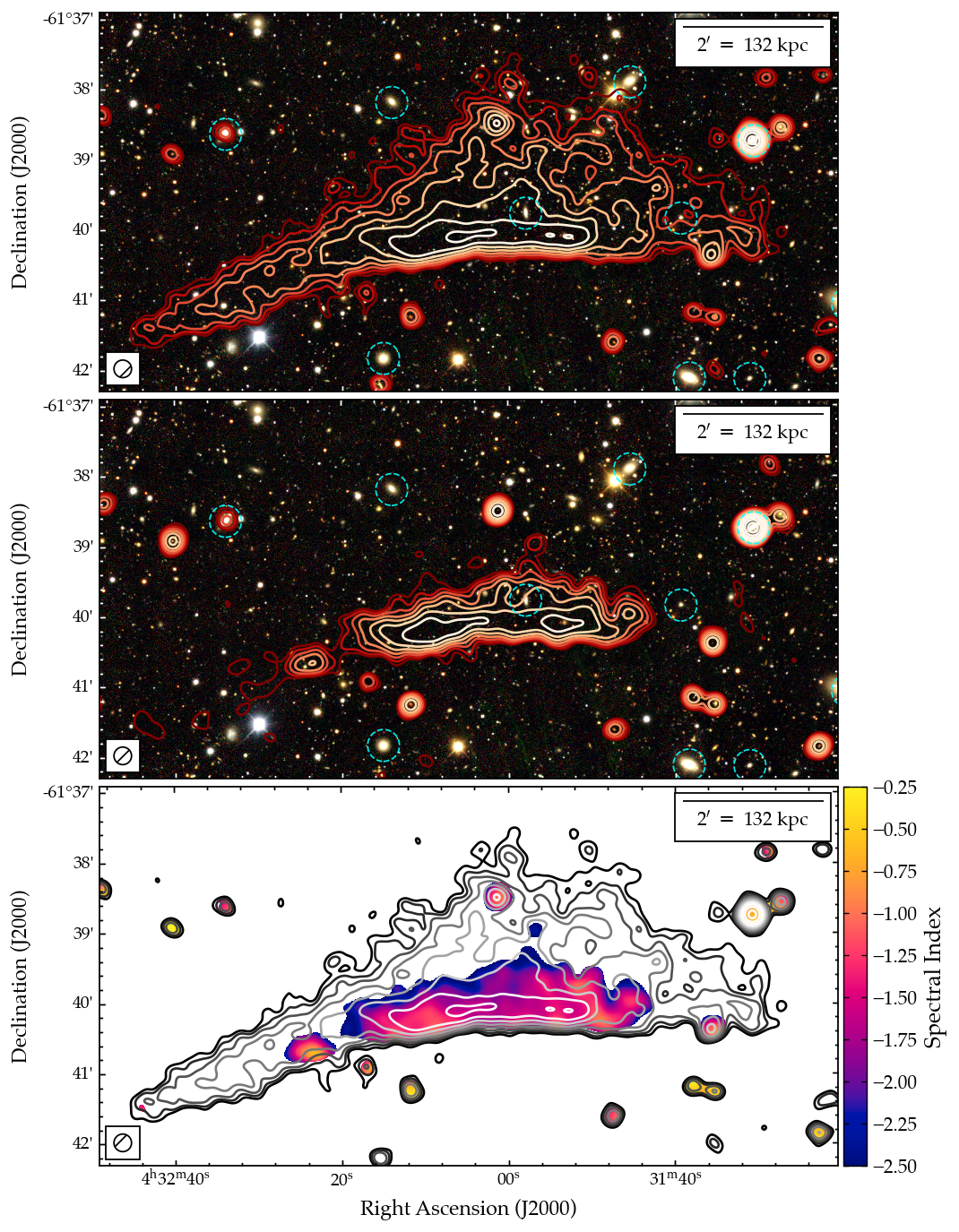}
\cprotect\caption{Radio-on-optical overlay of the `wrong-way' relic to the SE of Abell 3266. Contours show the surface brightness measured by ASKAP (943~MHz; \emph{upper panel}) and the ATCA (2.1~GHz; \emph{middle panel}) in our 15~arcsec images. Colormap shows our optical RGB image from DES, as per Figure~\ref{fig:A3266_RGB}. Dashed cyan circles denote cluster-member galaxies identified by \cite{Dehghan2017}. The \emph{lower panel} presents the spectral index map between 943~MHz and 2.1~GHz at 15~arcsec resolution. Contours show the ASKAP surface brightness. In each panel, contours start at $3\sigma$ and scale by a factor of $\sqrt{2}$, and the hatched circle in the lower-left corner indicates the restoring beam size (15~arcsec).}
\label{fig:WrongWayRelic_ThreePanel}
\end{center}
\end{figure}

This source lies at a projected distance of 15.8~arcmin ($\sim1.04$~Mpc) to the SE of the cluster centre. Source D1 is a highly-extended, asymmetric diffuse radio source --- in this case, a radio relic. This relic was tentatively identified by \cite{1999Murphy} at around $2.5\sigma$ significance; the first unambiguous detection was recently reported by \cite{Knowles2022}, and the low-frequency properties first explored by \cite{Duchesne2022}. Figure~\ref{fig:WrongWayRelic_ThreePanel} presents radio surface brightness contours of this source from our 15~arcsec resolution ASKAP (\emph{upper panel}) and ATCA (\emph{middle panel}) maps, overlaid on optical RGB image (as per Figure~\ref{fig:A3266_RGB}).

\subsubsection{Morphological properties}
From Figure~\ref{fig:WrongWayRelic_ThreePanel} this source exhibits a relatively bright, central, arc-like component with increasingly faint diffuse emission trailing North toward the cluster centre at 943~MHz. Faint `wings' of diffuse emission extend to the East and West of the central arc-like component. At 2.1~GHz, only the brighter central component is visible, implying that the wings and trailing emission are likely steep spectrum in nature.

We measure a largest angular size (LAS) of 8.7~arcmin at 943~MHz, which corresponds to a largest linear size (LLS) of 579~kpc. While this is smaller than many relics, this LLS is not unusual. We also highlight that the relic is slightly less extended at 2.1~GHz, where we measure a LAS $\sim6$~arcmin, or 394~kpc; this slight decrease in size with increasing frequency is also typical for relics.

At first glance, the morphology of this relic appears contradictory -- the concave shape appears to imply a merger shock travelling from the South to the North, whereas the brightness distribution visibly declining toward the cluster centre would imply a shock propagating from North to South. As such, this source has been dubbed a `wrong-way' relic\footnote{Priv. comm. Kenda Knowles.}. While radio relics with similar morphologies are uncommon, ours is not the only example: highly sensitive low-frequency observations have begun to reveal other concave relics \citep{HyeongHan2020,Botteon2021}.

Simulations suggest that highly-complex cluster merger events involving multiple sub-clusters can generate inhomogeneous shock structures that generate concave radio relics similar to that observed here \citep[][Wittor et al., in prep.]{Skillman2013}. As indicated by \cite{Dehghan2017}, Abell 3266 exhibits high velocity dispersion and significant sub-structure, implying a highly complex merger event, and so we suggest this source is likely a radio relic tracing a shock propagating in a N-S direction following the main cluster merger.

\subsubsection{Spectral properties}
The lower panel of Figure~\ref{fig:WrongWayRelic_ThreePanel} presents the spectral index map of the `wrong-way' relic between 943~MHz and 2.1~GHz. From Figure~\ref{fig:WrongWayRelic_ThreePanel}, the median spectral index is $\langle \alpha \rangle = -1.7$. Toward the south, at the presumed leading edge of the relic where particle acceleration is expected to occur due to the shock, the spectral index is marginally flatter with a median of $\langle \alpha \rangle = -1.4$. The spectral index steepens significantly toward the north in the presumed downstream region, reaching values around $\upalpha \sim -2$. The typical uncertainty in the spectral index map is about $0.1$ to $0.2$. This clear spectral gradient is typical of radio relics, as the synchrotron-emitting electrons age in the wake of the passing shock, which is located at the leading edge of the source. We note that the internal spectral indices from MeerKAT indicate steeper spectral indices for the relic than shown in Figure~\ref{fig:WrongWayRelic_ThreePanel}. However, since the robustness of these spectral indices as a function of distance from the pointing centre are still being studied, we adopt the current measurements for the purpose of our discussion.

While the LAS of the relic is around half the theoretical LAS accessible by our ATCA data, to check whether we were missing any flux from our ATCA image due to scale size sensitivity, we repeated this analysis using images produced with matched \emph{uv}-coverage, applying an inner \emph{uv}-cut of $122\lambda$ to our ASKAP data. From our \emph{uv}-matched images, we find negligible difference in the integrated flux density recovered. As such, we are confident that our following analysis is not biased by any inconsistencies in angular scale sensitivity. However, in the downstream region, the faint steep-spectrum emission may be too faint to recover by the ATCA at 2.1~GHz. Much deeper observations would be required to probe the spectral index of the emission in this region.

\begin{figure}
\begin{center}
\includegraphics[width=0.95\linewidth]{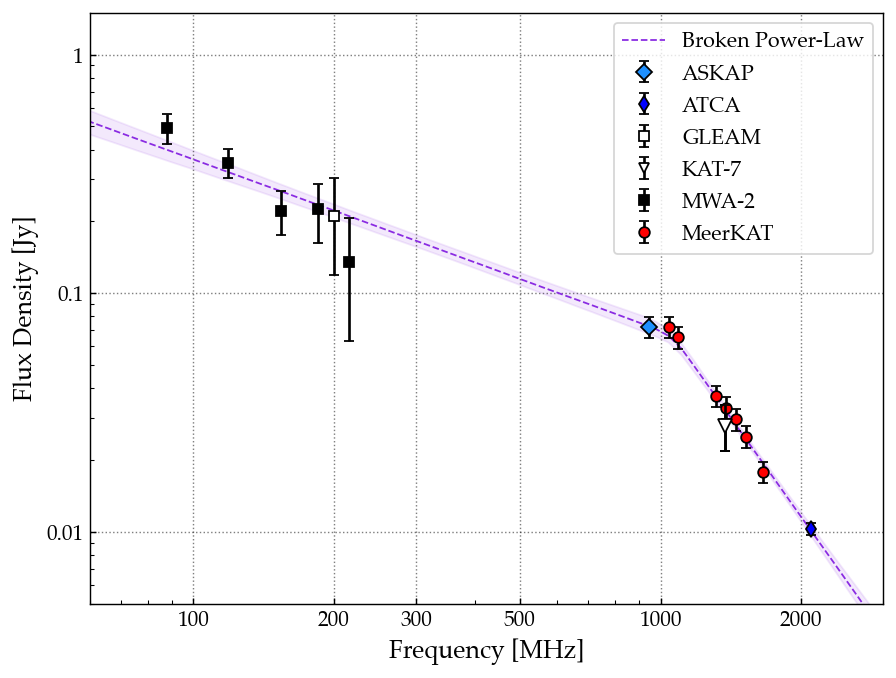}
\cprotect\caption{Integrated spectral energy distribution (SED) for the `wrong-way' relic to the SE of Abell 3266 (Source D1; shown in Figure~\ref{fig:WrongWayRelic_ThreePanel}). Our new ASKAP and ATCA integrated flux density measurements are shown as diamonds; our measurements from the public GCLS maps \citep{Knowles2022} are shown as red circles. We also report flux density measurements from the MWA and KAT-7 \citep{Duchesne2022}. The purple line shows our best-fit broken power-law spectrum; the shaded region denotes the $1\sigma$ uncertainty.}
\label{fig:Relic_SED}
\end{center}
\end{figure}

Integrating over the area defined by the $3\sigma$ contour of 15~arcsec ASKAP map presented in Figure~\ref{fig:WrongWayRelic_ThreePanel}, we measure a total flux density of $S_{\rm{943~MHz}} = 72.3 \pm 7.4$~mJy with ASKAP and $S_{\rm{2.1~GHz}} = 10.3 \pm 0.6$~mJy with the ATCA (after removing compact source contamination). To explore the broad-band SED of the `wrong-way' relic further, we integrated over the same area in the MeerKAT GCLS sub-band images selected for this work, as well as including flux density measurements from the MWA and KAT-7 reported by \cite{Duchesne2022}. Our new flux density measurements are reported as the `ASKAP-matched' measurements in Table~\ref{tab:relic_flux}. For consistency, we also integrated over the area of the relic defined in Figure~1 of \cite{Duchesne2022}. These larger-area measurements are consistent to within $1\sigma$ to $2\sigma$; we also report them as the `MWA-matched' measurements in Table~\ref{tab:relic_flux}. The broad-band integrated SED between 88~MHz and 2.1~GHz, derived using the ASKAP-matched measurements is presented in Figure~\ref{fig:Relic_SED}.

\begin{table}
\footnotesize
\centering
\caption{Integrated flux density measurements for the `wrong-way' relic in Abell~3266. The `ASKAP-matched' column reports the measurement integrated over the area defined by the $3\sigma$ contour in the top panel of Figure~\ref{fig:WrongWayRelic_ThreePanel}; the `MWA-matched' column reports the measurement integrated over the full extent of the relic measured in Figure~1 of \protect\cite{Duchesne2022}. In all cases, contaminating compact sources have been excised. \label{tab:relic_flux}}
\begin{tabular}{lrccc}
\hline
            &           & \multicolumn{2}{c}{Integrated flux density} \\
Instrument  & Frequency & ASKAP-matched & MWA-matched \\
            & [MHz]     & [mJy] & [mJy] \\

\hline 
ASKAP       & 943       & $72.3 \pm 7.4$ & $77.3 \pm 7.8$ \\
MeerKAT     & 1043      & $72.1 \pm 7.4$ & $73.8 \pm 7.4$ \\
MeerKAT     & 1092      & $65.4 \pm 6.7$ & $67.8 \pm 6.9$ \\
MeerKAT     & 1312      & $37.2 \pm 3.8$ & $39.3 \pm 4.0$ \\
MeerKAT     & 1381      & $33.2 \pm 3.4$ & $34.7 \pm 3.5$ \\
MeerKAT     & 1448      & $29.7 \pm 3.1$ & $30.3 \pm 3.1$ \\
MeerKAT     & 1520      & $25.1 \pm 2.6$ & $28.7 \pm 2.9$ \\
MeerKAT     & 1656      & $17.8 \pm 1.8$ & $20.6 \pm 2.1$ \\
ATCA        & 2100      & $10.3 \pm 0.6$ & $11.8 \pm 0.7$ \\
\hline
\end{tabular}
\end{table}

From Figure~\ref{fig:Relic_SED}, the `wrong-way' relic exhibits clear departure from a single power-law behaviour, atypical for relics associated with merger shocks. To the best of our knowledge, only two relics in the literature have previously been reported with broken power-law spectra: the relic in Abell~2256 \citep{Owen2014,Trasatti2015} and the North Relic in ZwCl~2341.1$+$0000 \citep{Parekh2022}. However, recent work has shown that the relic in Abell~2256 in fact shows single power-law behaviour up to at least 3~GHz \citep{Rajpurohit2022}. Furthermore, the SED for the North Relic in ZwCl~2341.1$+$0000 presented by \cite{Parekh2022} shows measurements at only four frequencies, so while the SED steepens it does not follow a clearly different power-law at higher frequencies. With the addition of JVLA data in the 2$-$4~GHz range originally reported by \cite{Benson2017}, both relics in ZwCl~2341.1$+$0000 appear to follow a single power law across the frequency range 610~MHz to 3~GHz.

Our analysis involves broad-band data across a much wider frequency range, with clear --- and distinctly different --- power-law behaviour at both high and low frequencies. As such, the `wrong-way' relic in Abell~3266 is arguably the clearest example of a relic which shows significant spectral steepening reported to-date.

To quantify the spectral steepening we approximate both the low-frequency part ($\nu \leq 943$~MHz) and the high-frequency part ($\nu \geq 943$~MHz) with a power-law spectrum. We used an `affine invariant' Markov-chain Monte Carlo (MCMC) ensemble sampler \citep{Goodman2010} as implemented by the \textsc{emcee} package \citep{emcee} to constrain the model parameters of this broken power-law fit. The best-fit SED model, along with the 16th and 84th percentile uncertainty region, is plotted in Figure~\ref{fig:Relic_SED}. Our modelling yields a low-frequency spectral index $\alpha_{\rm low} = -0.72 \pm 0.06$ below the break frequency ($\nu_{\rm b} = 1056 \pm 49$~MHz) and $\alpha_{\rm high} = -2.76 \pm 0.15$ above this frequency. This high-frequency spectral index is extremely steep for radio relics.

\begin{figure*}
\begin{center}
\includegraphics[width=0.99\linewidth]{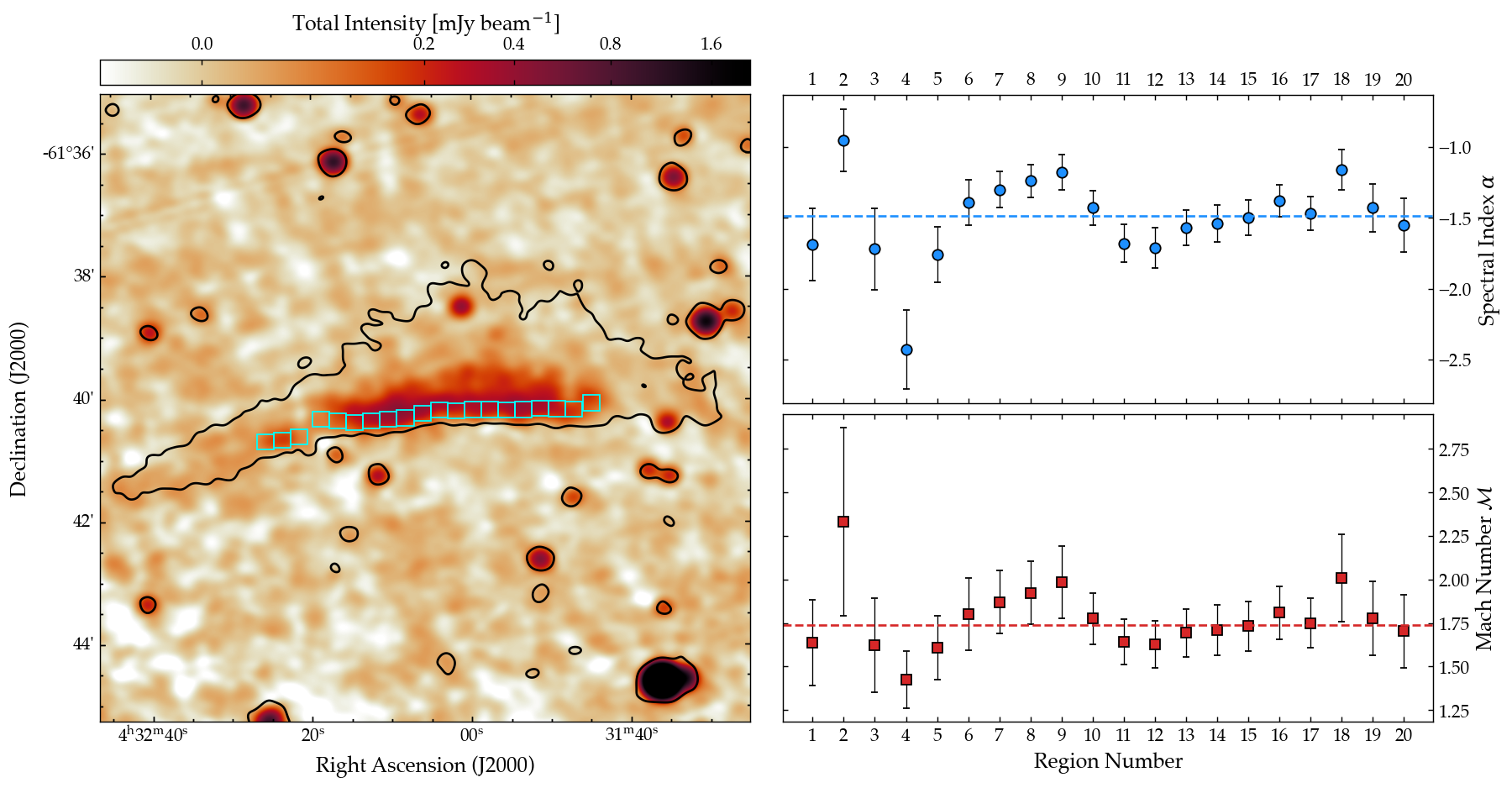}
\cprotect\caption{\emph{Left panel:} radio surface brightness maps of the `wrong-way' relic at 15~arcsec resolution. Colormap shows our ATCA image at 2.1~GHz; black contour shows the $4\sigma$ contour from our ASKAP map. Cyan boxes show 15~arcsec regions used to profile the spectral index and Mach number. Box numbers increase monotonically from left to right. \emph{Right panels:} profile of the spectral index $\alpha$ (\emph{upper}, in blue) and Mach number $\mathcal{M}$ (\emph{lower}, in red) for regions shown in the left panel. Dashed lines show the median values of the spectral index, $\langle \alpha \rangle = -1.48 \pm 0.14$, and Mach number, $\langle \mathcal{M} \rangle = 1.74 \pm 0.23$, for the respective subplot.}
\label{fig:WrongWayRelic_Profile}
\end{center}
\end{figure*}

Given the spectral index convention we adopt, the $k$-corrected radio power $P_{\nu}$ at frequency $\nu$ is given by the following equation:
\begin{equation}\label{eq:radio_lum}
    P_{\nu} = 4 {\mathrm{\pi}} \,  D_{\rm L}^2 \, S_{\nu} \, (1 + z)^{-(1 + \alpha)}
\end{equation}
where $D_{\rm L} = 255$~Mpc is the luminosity distance at the cluster redshift of $z = 0.0594$, and $S_{\nu}$ is the flux density at frequency $\nu$. Based on our broad-band SED fit presented in Figure~\ref{fig:Relic_SED}, we use Equation~\ref{eq:radio_lum} to derive a radio power of $P_{\rm{1.4~GHz}} = (2.38 \pm 0.10) \times 10^{23}$~W~Hz$^{-1}$. While this places the `wrong-way' relic in Abell~3266 toward the low-power end of the known population, it is not visibly an outlier in typical scaling relation planes between radio relic power and cluster mass, for example \citep{Duchesne2021a,Duchesne2021d}.

\subsubsection{Acceleration mechanisms}
The commonly-accepted theoretical framework for generating radio relics is DSA, whereby a population of electrons are accelerated to the relativistic regime by merger shocks. These shocks are often detectable at X-ray wavelengths, presenting as a jump in the ICM density, pressure, and/or temperature. While our \xmm{} mosaic covers a large area, including deep integrations on both the cluster centre as well as the outskirts, there is insufficient signal-to-noise to perform a meaningful investigation into the ICM properties at the location of the `wrong-way' relic. However, \cite{Sanders2022} report the presence of a shock at a distance of 1.1~Mpc from the cluster centre measured with eROSITA, consistent with the position of the relic. Depending on whether the pressure jump or density jump is used to trace the discontinuity, \citeauthor{Sanders2022} find a Mach number $\mathcal{M_{\rm X-ray}} \sim 1.54 - 1.71$.

Cluster merger shocks in the ICM can be considered as quasi-stationary if the timescale on which the geometry and/or shock strength changes is much greater than the cooling timescale of the synchrotron-emitting electrons. Under this stationary shock situation, the integrated spectral index $\upalpha_{\rm{int}}$ is slightly steeper than the injection index $\upalpha_{\rm{inj}}$, following:
\begin{equation}\label{eq:inj_index}
    \upalpha_{\rm{int}} = \upalpha_{\rm{inj}} - 0.5
\end{equation}

Assuming the underlying physical mechanism responsible for generating the relic is a form of DSA, the relationship between the radio-derived shock Mach number and the injection index takes the following form:
\begin{equation}\label{eq:mach_radio}
    \mathcal{M}_{\rm radio} = \left( \frac{2\upalpha_{\rm{inj}} - 3}{2 \upalpha_{\rm{inj}} + 1} \right)^{\frac{1}{2}}
\end{equation}
originally presented by \cite{BlandfordEichler1987}.

However, as shown in Figure~\ref{fig:Relic_SED}, the integrated spectrum of the `wrong-way' relic shows significant steepening. Thus, if the `wrong-way' relic is caused by a shock front which propagates into the cluster outskirts, the shock front may not cause radio emission according to the quasi-stationary shock scenario, and simply determining the Mach number using the \emph{integrated spectral index} is invalid; instead we use the \emph{observed spectral index} at the shock front to represent the injection index in Equation~\ref{eq:mach_radio}. To better explore the sub-structure at the shock front, we profiled our spectral index map using adjacent boxes of 15~arcsec width, spaced across the long axis of the relic. The regions used, along with the resulting spectral index and Mach number profile, are presented in Figure~\ref{fig:WrongWayRelic_Profile}.

From Figure~\ref{fig:WrongWayRelic_Profile}, we find a median spectral index of $\langle \alpha \rangle = -1.48 \pm 0.14$ at the leading edge of the relic. We also see that the spectral index profile of the `wrong-way' relic varies somewhat across its length, though is typically in the range  $-1.75 \lesssim \alpha \lesssim -1.25$ across the majority of regions. Several regions toward the easternmost edge of the relic show significantly varying spectral behaviour. While the signal-to-noise ratio is lower at these extremes (and thus the uncertainty is greater), the divergence from the median value is significant, perhaps indicating different physical conditions. 

The Mach number profile in the lower-right panel of Figure~\ref{fig:WrongWayRelic_Profile} indicates that the `wrong-way' relic traces a distribution of weak Mach numbers $1.4 \lesssim \mathcal{M} \lesssim 2.3$, with a median value of $\langle \mathcal{M} \rangle = 1.74 \pm 0.23$. As we are measuring the variations in spectral index locally, we are resolving spectral variations across the leading edge of the relic (and hence variations in the strength of the shock). Simulations by \cite{Wittor2021} for example show that the Mach number distribution generated by cluster mergers is broader, with a typically lower average Mach number, than is inferred from the integrated spectral index (which traces the \emph{radio-weighted Mach number}). Such behaviour is also seen in some other well studied relics \citep[e.g.][]{DiGennaro2018,Rajpurohit2018,Rajpurohit2021c}.

Some previous studies show tension between radio- and X-ray-derived Mach numbers \citep[e.g.][]{Akamatsu2017,Urdampilleta2018} and while it may indicate underlying problems with our understanding of the DSA mechanism, recent simulations have gone some way to reconciling the discrepancy. \cite{Wittor2021} showed that the radio emission from relics traces a \emph{distribution} of Mach numbers, and is hence sensitive to high Mach number shocks at the tail of the distribution, whereas the X-ray emission provides a better tracer of the \emph{average} of the Mach number distribution. Additionally, \cite{Dominguez-Fernandez2021} showed that the properties of radio relics are heavily influenced by the properties of the ICM, and that fluctuations in the ICM (due to turbulence, for example) also act to reconcile the tension between $\mathcal{M_{\rm X-ray}}$ and $\mathcal{M_{\rm radio}}$.

Overall, our radio-derived Mach number would appear to be consistent with the X-ray-derived Mach number. However, we note that given the clear spectral steepening in the SED, the injection index cannot be inferred from this radio Mach number. The low radio Mach number we have derived from the spectral index would result in a relic with an observed luminosity based on the classical DSA scenario for radio relics, which clearly cannot be the case.

\subsubsection{On the nature of the `wrong-way' relic}
The `wrong-way' relic in Abell~3266 is a highly atypical example of its kind. To summarise our findings: (i) its morphology is somewhat unusual, although it is not the only example of a concave relic in the literature \citep[e.g.][]{HyeongHan2020,Botteon2021}; (ii) it coincides with an X-ray shock that exhibits a low Mach number; (iii) it shows clear spectral steepening in the downstream region; (iv) it exhibits significant steepening in the integrated spectrum, with a high-frequency spectral index that is extremely steep for a radio relic, and a low-frequency spectral index that is flatter than is expected for an injection-plus-ageing electron population generated by the standard quasi-stationary shock scenario (which would require a spectral index steeper than $-1$). The clear downstream spectral index trend favours the classification as radio relic and disfavours alternative explanations (such as a radio phoenix). The `wrong way' morphology is exceptional since it combines evidence for being caused by the shock front with an extreme spectral steepening. It is challenging to find a plausible scenario for the extreme steepening.

One of the big open questions in studies of DSA in radio relics is whether the shock is efficient enough to accelerate electrons from the thermal pool to the relativistic regime, or whether a pre-existing population of mildly-relativistic electrons is required. Such a population may have been injected over the course of a cluster's lifetime by radio galaxy jets, for example. The majority of the more powerful radio relics appear to challenge standard DSA from the thermal pool, requiring unphysically-large acceleration efficiency to generate the observed high radio luminosity \citep[e.g.][]{Kang2011,Botteon2016a,Botteon2020a}. However, some radio relics appear to be consistent with acceleration directly from the thermal pool \citep[e.g.][]{Botteon2016b,Botteon2020a,Locatelli2020}. 

In the case of the `wrong-way' relic, the clear spectral steepening seen in the integrated spectrum (Figure~\ref{fig:Relic_SED}) provides strong evidence of a more complex scenario than standard DSA. Alternatives to the standard DSA scenario have been proposed, such as the shock \emph{re}-acceleration of a pre-existing mildly-relativistic electron population \citep[e.g.][]{Kang2011,Kang2016} or an electron population that experiences multiple shocks \citep{Inchingolo2022}. It is of particular interest to note the morphological and spectral similarities between the `wrong-way' relic in Abell~3266 and `Relic B' in the simulations of \cite{Inchingolo2022}, which also shows a somewhat concave morphology.

However, both of these alternative scenarios generate integrated spectra that are \emph{curved} rather than showing the striking broken power-law seen in Figure~\ref{fig:Relic_SED}. To the best of our knowledge, there is currently no physically-motivated theoretical framework that can generate this spectrum. This motivates deeper theoretical work to determine the exact physical conditions required to yield broken power-law behaviour seen in Figure~\ref{fig:Relic_SED}. One potential avenue for investigation would be to examine whether alternative scenarios could lead to a significantly curved spectrum at injection. However, such work is beyond the scope of this paper.

\subsection{Source D2: an ultra-steep spectrum fossil}
\begin{figure}
\begin{center}
\includegraphics[width=0.95\linewidth]{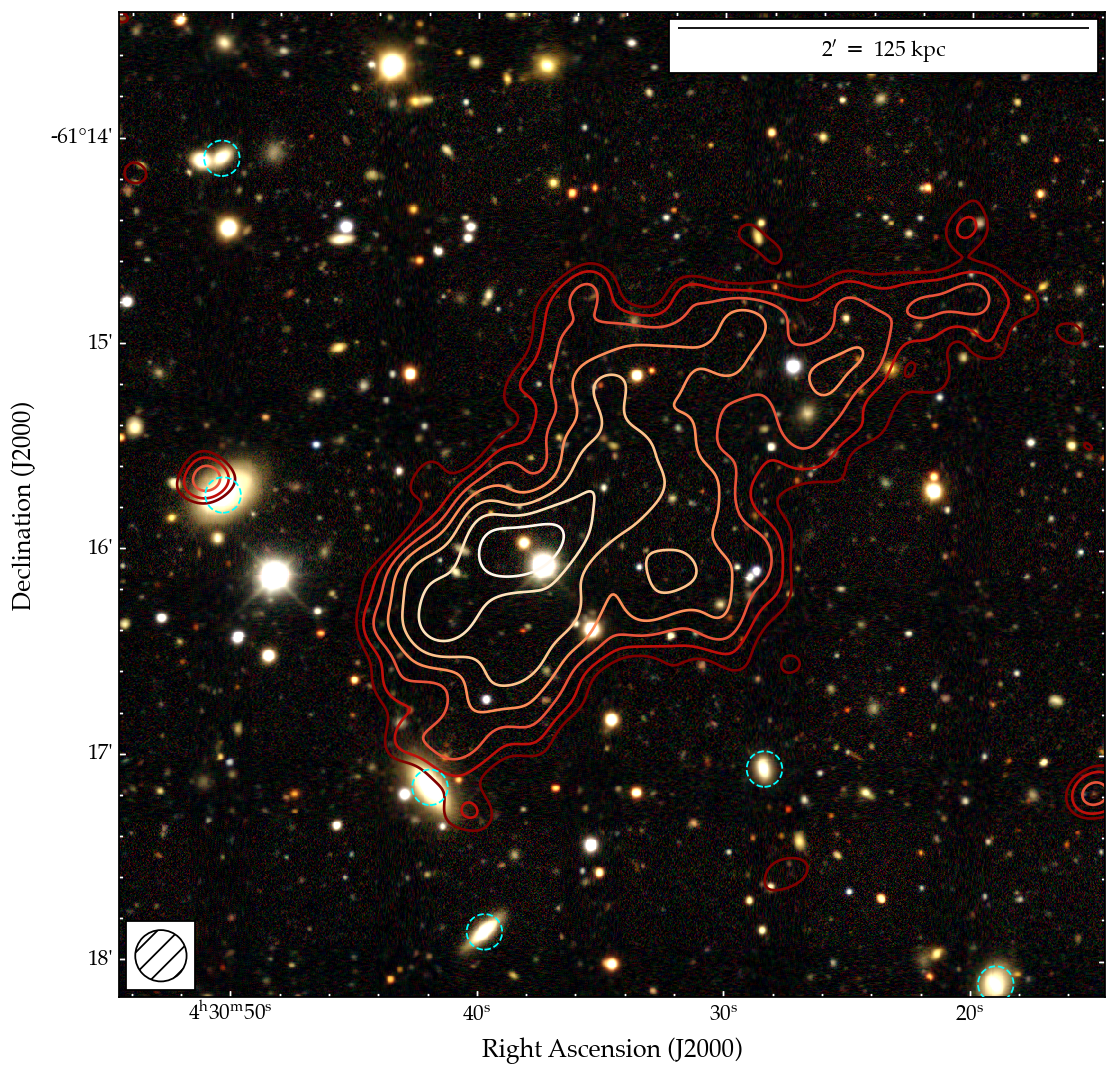}
\cprotect\caption{Radio-on-optical overlay of the ultra-steep-spectrum fossil radio galaxy to the NW of Abell 3266 (source `D2' in Figure~\ref{fig:A3266_RGB}). Contours show the 943~MHz ASKAP surface brightness at 15~arcsec resolution, starting at $3\sigma$ and scaling by a factor of $\sqrt{2}$. Colormap shows an optical RGB image from DES, as per Figure~\ref{fig:A3266_RGB}. Dashed cyan circles denote cluster-member galaxies identified by \cite{Dehghan2017}. The proposed host for this diffuse radio source is the cluster-member galaxy at the SE tip of the emission.}
\label{fig:Fossil_Overlay}
\end{center}
\end{figure}

This source, first reported by \cite{Duchesne2022}, lies to the NW of the cluster centre. It is asymmetric and moderately extended at 943~MHz, with a LAS of 3.6~arcmin. It is undetected by the ATCA at 2.1~GHz. Figure~\ref{fig:Fossil_Overlay} presents the ASKAP surface brightness contours of this source, overlaid on our RGB image from DES (as per Figure~\ref{fig:A3266_RGB}). 

Two possible optical counterparts are visible in Figure~\ref{fig:Fossil_Overlay}. One aligns roughly with the radio centroid, but cannot be the host as it is a foreground star, so we suggest that the cluster-member galaxy toward the SE tip of the emission (identified by \citealt{Dehghan2017}) was the likely original host of this diffuse radio source. The cluster-member galaxy J043041.92$-$611709.7 has a redshift $z = 0.05587$ and is a member of Structure~5, one of the sub-groups surrounding the core of Abell 3266 \citep[see][]{Dehghan2017}. This suggests a projected LLS of 223~kpc.

The diffuse, highly-asymmetric nature of this object, combined with the lack of an identifiable core at either 943~MHz or 2.1~GHz, supports the suggestion by \cite{Duchesne2022} that this is a fossil source originally associated with J043041.92$-$611709.7. Following this scenario, the fossil radio jets extend out from Structure~5 into the region between Structures~3 and 4, where no cluster-member galaxies have been identified. There are a variety of assessment criteria for selecting candidate remnant radio galaxies, including morphological properties, ultra-steep (and curved) spectral index, and core prominence, that have been used in numerous previous remnant studies \citep{Brienza2017,Mahatma2018,Jurlin2021,Quici2021}. The morphological properties are consistent with the proposed remnant scenario; we will now consider the spectral properties and core prominence.

\subsubsection{Spectral properties}
ASKAP measures a peak flux density of $S_{\rm{pk,~943~MHz}} = 1.48$~mJy beam$^{-1}$ in our 15~arcsec resolution image. Despite achieving an rms of $13.9~\upmu$Jy beam$^{-1}$ with the ATCA, this source remains undetected at 2.1~GHz. As such, taking a conservative $2\sigma$ upper limit, we can place an upper limit of $\upalpha_{\rm 943~MHz}^{\rm 2.1~GHz} \leq -4.9$ on the spectral index of this object in the frequency range 943~MHz to 2.1~GHz. This is an extremely steep spectrum, approaching the steepest spectral index measured to-date \citep{Hodgson2021}, although the spectral index measured by \citeauthor{Hodgson2021} is at much lower frequency, implying an older population of electrons.

At much lower resolution, this ultra-steep-spectrum diffuse object in Abell~3266 has been previously catalogued as part of the GaLactic and Extragalactic All-sky MWA survey \citep[GLEAM;][]{Wayth2015}. The GLEAM Extragalactic Catalogue \citep{HurleyWalker2017} object is GLEAM~J043032$-$611544; it is an unresolved source with a 200~MHz flux density of $S_{\rm{int,~200~MHz}} = 1.45 \pm 0.14$~Jy. Despite comparable sensitivity and resolution, it is also undetected by KAT-7, supporting the suggestion of an extremely steep spectral index. \cite{Duchesne2022} present images of this cluster from the Phase II MWA; they find that this source has a steep low-frequency spectral index $\alpha_{\rm 88~MHz}^{\rm 216~MHz} = -1.70 \pm 0.14$. This steep spectral index implies an aged plasma, even at low radio frequencies.

\begin{figure}
\begin{center}
\includegraphics[width=0.95\linewidth]{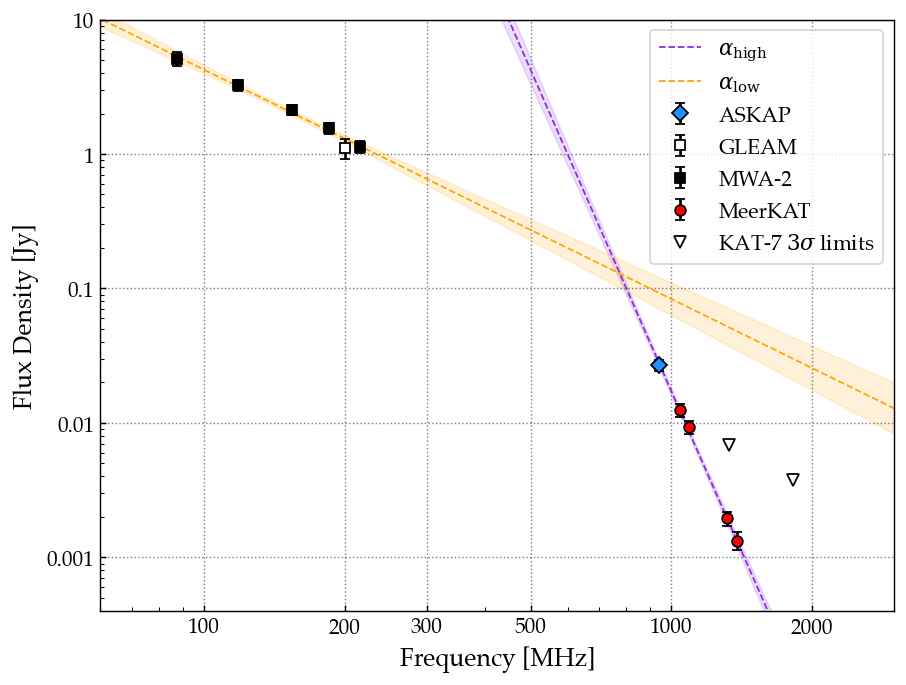}
\cprotect\caption{Spectral energy distribution (SED) for the ultra-steep spectrum fossil to the NW of Abell 3266 (Source D2; shown in Figure~\ref{fig:Fossil_Overlay}). Our new ASKAP measurement is shown by the blue diamond; datapoints from the MWA \citep{Duchesne2022} are shown as squares. Open triangles denote $3\sigma$ upper limits from KAT-7. In-band spectral measurements from the MeerKAT GCLS data (where the source was detectable above $3\sigma$) are shown in red. The spectral index is extremely steep, approximated by a power-law fit of $\alpha_{\rm low} = -1.70 \pm 0.14$ between 88 and 216~MHz and $\alpha_{\rm high} = -7.92 \pm 0.30$ between 943 and 1312~MHz. These fits are shown as dashed lines, with the shaded region denoting the $1\sigma$ uncertainty.}
\label{fig:Fossil_SED}
\end{center}
\end{figure}

Integrating above the $3\sigma$ contour, ASKAP measures an integrated flux density of $S_{\rm{int,~943~MHz}} = 26.8 \pm 2.7$~mJy. We use this measurement in conjunction with the low-frequency Phase II MWA measurements presented by \cite{Duchesne2022}, plus measurements made on the MeerKAT GCLS data products, to explore the broad-band integrated SED for this fossil source. As for the `wrong-way' relic, we report two sets of measurements: an `ASKAP-matched' flux density, integrated over the area defined by the $3\sigma$ contour in Figure~\ref{fig:Fossil_Overlay}, and an `MWA-matched' flux density, integrated over the area defined in Figure~1 of \cite{Duchesne2022}. These larger-area measurements are also broadly consistent with the ASKAP-matched measurements. Our measurements are listed in Table~\ref{tab:fossil_flux} and the integrated SED for Source D2 (using our `ASKAP-matched' measurements) is presented in Figure~\ref{fig:Fossil_SED}.

\begin{table}
\footnotesize
\centering
\caption{Integrated flux density measurements for source D2, the ultra-steep spectrum fossil to the NW of Abell~3266. The `ASKAP-matched' column reports the measurement integrated over the area defined by the $3\sigma$ contour of Figure~\ref{fig:Fossil_Overlay}. The `MWA-matched' column reports the measurement integrated over the full extent of the fossil measured in Figure~1 of \protect\cite{Duchesne2022}, after subtracting compact sources within the region. This source becomes undetectable in the MeerKAT GCLS maps above 1381~MHz due to its ultra-steep spectrum. \label{tab:fossil_flux}}
\begin{tabular}{lrrr}
\hline
            &           & \multicolumn{2}{c}{Integrated flux density} \\
Instrument  & Frequency & ASKAP-matched & MWA-matched \\
            & [MHz]     & [mJy] & [mJy] \\
\hline 
ASKAP       & 943       & $26.9 \pm 2.7$    & $27.8 \pm 3.0$  \\
MeerKAT     & 1043      & $12.5 \pm 1.3$    & $12.5 \pm 1.3$  \\
MeerKAT     & 1092      & $9.4 \pm 1.0$     & $9.5 \pm 1.0$  \\
MeerKAT     & 1312      & $1.9 \pm 0.2$     & $2.4 \pm 0.3$  \\
MeerKAT     & 1381      & $1.3 \pm 0.2$     & $1.6 \pm 0.2$  \\
\hline
\end{tabular}
\end{table}

It is immediately striking that the SED is highly curved, and the deeper data provided by ASKAP and MeerKAT are consistent with the upper limits from KAT-7 \citep[][]{RiseleyThesis,Duchesne2022}. The spectra of remnants are often described using models such as the `JP' model \citep{JaffePerola1973}, the `KP' model \citep{Kardashev1962,Pacholczyk1970}, or a form of the `continuous injection' or `CI' model \citep{Komissarov1994}. The `CI' model describes the spectrum arising from an electron population injected over a period from $t = 0$ to $t = \tau$, where $\tau$ is the source age. Two versions of the `CI' model exist: `CI-on', where energy injection is still occurring, and `CI-off', where the injection ceased some period of time ago ($\tau_{\rm off}$). These models all predict a spectrum that becomes increasingly steep and curved as a source ages, with spectral break that moves to increasingly lower frequencies. We attempted to fit `JP', `KP' and `CI-off' models using the \textsc{synchrofit} package \citep[developed as part of][]{Quici2022}\footnote{Currently available at \url{https://github.com/synchrofit/synchrofit}}. None provided a good description of the SED behaviour shown in Figure~\ref{fig:Fossil_SED}.

This is unsurprising: the ultra-steep low-frequency spectral index $\alpha_{\rm 88~MHz}^{\rm 216~MHz} = -1.70 \pm 0.14$ measured by \cite{Duchesne2022} implies an injection index that is steeper than is expected from theory, and cannot be replicated by any of the aforementioned models. We approximated the high-frequency spectral index using a power-law fit to the measurements between 943 and 1312~MHz from Table~\ref{tab:fossil_flux}, finding a value of $\alpha_{\rm high} = -7.92 \pm 0.30$. This is extremely steep, and is not replicated by any of the physically-motivated models. Overall, it is likely that the spectral break for source D2 occurs below the observing band of the MWA, similar to the ultra-steep spectrum `jellyfish' source reported by \cite{Hodgson2021}.

From Equation~\ref{eq:radio_lum}, we can estimate the $k$-corrected 1.4~GHz radio luminosity. We must make some assumptions however; we assume that (i) our identification of the host galaxy is correct \citep[$z = 0.05587$;][]{Dehghan2017} and (ii) the spectral index is reasonably approximated by the ultra-steep power-law fit to our ASKAP and MeerKAT measurements. While the flux density of source D2 drops precipitously at these frequencies, and as such our modelling likely over-estimates the 1.4~GHz flux density, we estimate an upper limit to the 1.4~GHz flux density of $S_{\rm 1.4~GHz} = 1.35 \pm 0.14$~mJy. Equation~\ref{eq:radio_lum} yields an upper limit radio luminosity of $P_{\rm 1.4~GHz} = (1.36 \pm 0.13) \times 10^{22}$~W~Hz$^{-1}$.

\subsubsection{Core prominence}\label{sec:fossil_cp}
The core prominence $(\rm CP)$ quantifies the brightness of the radio core, compared to the total brightness of a radio galaxy. It is defined as the ratio of core flux to total flux, i.e. $\rm CP = S_{\rm core} / S_{\rm int}$. Typically, remnant radio galaxies compiled from historic samples of powerful radio galaxies have a core prominence $\rm CP \lesssim 10^{-4}$ \citep{Giovannini1988}; however, more recent, highly sensitive studies suggest that aged remnants may be preferentially deselected if a low core prominence is used as a criterion \citep{Brienza2017,Quici2021}.

In the case of our fossil, Source D2, no core is visible at either 943~MHz or 2.1~GHz. We therefore place a conservative upper-limit on $S_{\rm core}$ by measuring the local rms noise at the location of the proposed host galaxy, and taking $S_{\rm core, 943~MHz} = 3\sigma_{\rm host} \simeq 0.246$~mJy. The core prominence is therefore $\rm CP \lesssim 10^{-2}$. Thus, this source would not be selected as a remnant \emph{solely} on the basis of CP, consistent with recent statistical studies of AGN remnant populations \citep{Brienza2017,Quici2021}. Overall, given the radio power derived above, source D2 lies some way below the plane of the $\rm CP / P_{\rm 1.4~GHz}$ relation for remnant AGN \citep[e.g.][]{DeRuiter1990,Jurlin2021}, and is consistent with the USS remnant population modelled by \cite{Brienza2017} for Lockman Hole field, when taking into account both radiative and dynamical evolution.

\subsection{Source D3: the central diffuse ridge}
\begin{figure*}
\begin{center}
\includegraphics[width=0.99\textwidth]{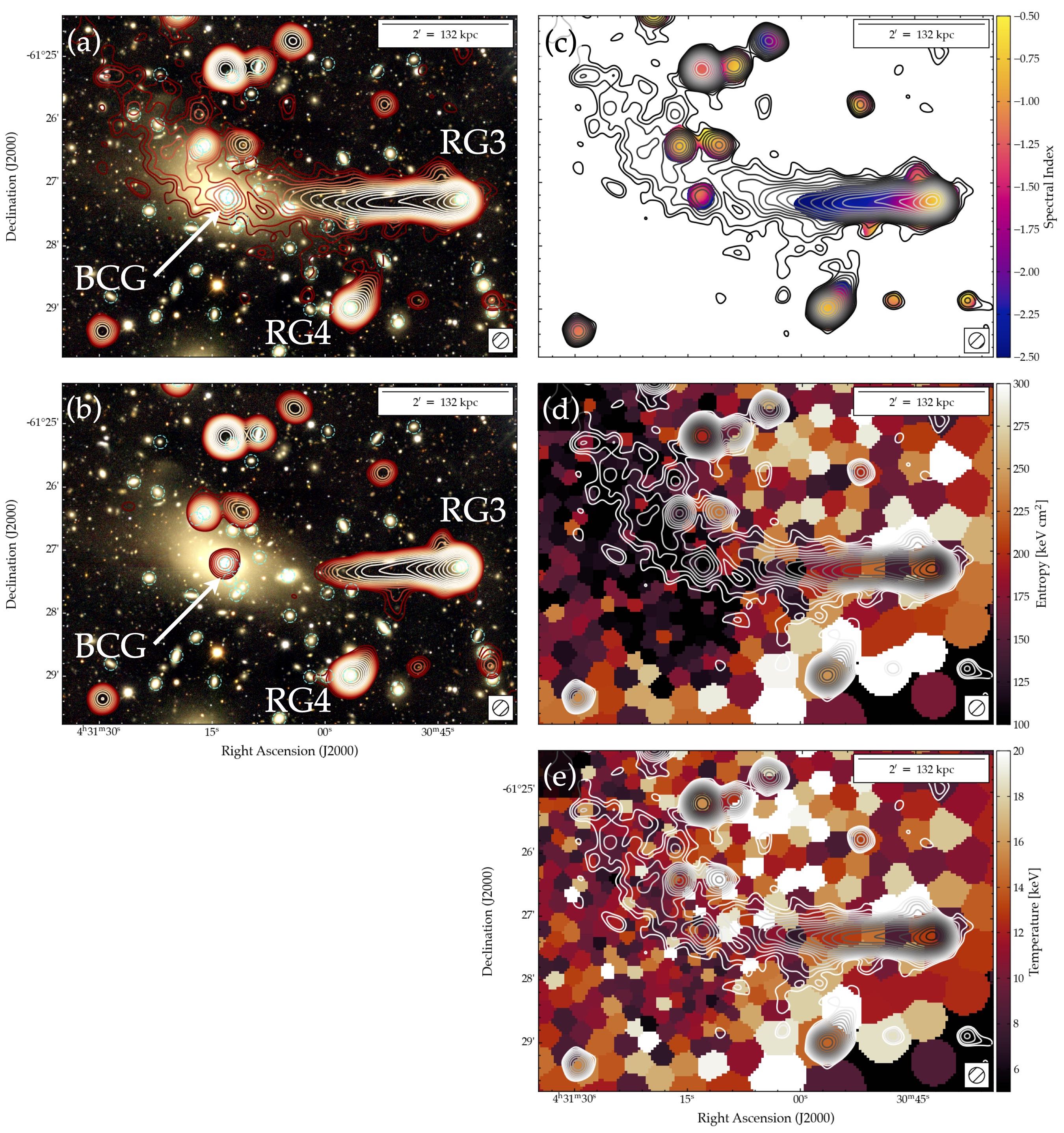}
\cprotect\caption{Zoom on the central region of Abell~3266. Panels (a) and (b) show our colour-composite DES image with radio contours from ASKAP at 943~MHz and the ATCA at 2.1~GHz, respectively. Cyan circles identify known cluster-member galaxies. We also highlight two radio galaxies of interest, `RG3' and `RG4' as per Figure~\ref{fig:A3266_RGB}, as well as the BCG. Panel (c) shows the radio spectral index, derived using our ASKAP and ATCA data. Panels (d) and (e) show the pseudo-entropy and ICM temperature derived from our \xmm{} mosaic. Contours in panels (c), (d) and (e) show the radio surface brightness measured by ASKAP. In all panels, contours are shown at 15~arcsec resolution (indicated by the hatched circle in the lower-right corner) and scale by a factor of $\sqrt{2}$ from the $3\sigma$ level.}
\label{fig:central}
\end{center}
\end{figure*}

We present a zoom on the central region of Abell~3266 in Figure~\ref{fig:central}. From our ASKAP map at 15~arcsec resolution, we recover low-level diffuse emission -- a diffuse `ridge' -- associated with the central region of Abell~3266. The LAS of this emission is $218$~arcsec, corresponding to a LLS of 240~kpc. Integrating above the $3\sigma$ contour and excising the contribution from the three largely-compact radio galaxies embedded in this diffuse emission (including the BCG), ASKAP recovers a total integrated flux density of $S_{\rm 943~MHz} = 7.95 \pm 0.78$~mJy.

This emission is not detected in our ATCA maps at 2.1~GHz, suggesting a steep spectral index; the typical 943~MHz surface brightness is $\sim 150 - 250~\upmu$Jy beam$^{-1}$ (around $0.8~\upmu$Jy arcsec$^{-2}$). From our 15~arcsec resolution ATCA map, we measure a typical local rms of $13.1~\upmu$Jy beam$^{-1}$ (around $0.05~\upmu$Jy arcsec$^{-2}$); assuming a $2\sigma$ upper limit, we estimate a spectral index of $\alpha_{\rm 943~MHz}^{\rm 2.1~GHz} \lesssim -2.54$. Such an ultra-steep spectral index would be consistent with a highly inefficient acceleration process such as turbulent acceleration.

This emission is recovered at much lower spatial resolution in the MWA Phase II maps presented by \cite{Duchesne2022}, although the resolution is too coarse to make a more accurate determination of the spectral index. Similarly, in the MeerKAT GCLS maps of Abell~3266, this emission is only detected at frequencies below around 1100~MHz, which overlaps heavily with our ASKAP coverage. Given the limited dynamic range of the MeerKAT GCLS maps of Abell~3266 at these frequencies, as well as the narrow lever arm in frequency, any spectral index measurement made between ASKAP and MeerKAT would be highly uncertain.

Given that Abell 3266 is a high-mass cluster undergoing a complex merger event, we might expect the presence of a radio halo. However, the MWA Phase II data presented by \cite{Duchesne2022} show no evidence of a giant radio halo. Using a range of injected radio halo profiles (derived using the scaling relations presented in \citealt{Duchesne2021b}), \citeauthor{Duchesne2022} place an upper limit of $P_{\rm 1.4~GHz} \lesssim 0.3 - 0.8 \times10^{24}$~W~Hz$^{-1}$ on the luminosity of any halo that may be present, but below the detection limit of the Phase II MWA data. This is around a factor five below expectations from the scaling plane between cluster mass and radio halo power.

The upper limit to the spectral index implied by our ASKAP and ATCA maps is also far steeper than the typical spectral index of many radio haloes \citep[e.g.][]{Dallacasa2009,Knowles2016,Shweta2020,Wilber2020,Bruno2021,Duchesne2021c,Duchesne2021d,Rajpurohit2021b,Rajpurohit2021c}. Thus, the nature of this source is not immediately obvious. We explore two potential scenarios for the nature of this source: (i) some form of revived fossil plasma source that originated from one (or more) of the cluster-member radio galaxies, and (ii) a newly detected mini-halo.

\subsubsection{Association with cluster-member radio galaxies}
In this scenario, the diffuse central ridge could represent fossil plasma from nearby tailed radio galaxies that has been re-accelerated by turbulence in the wake of the cluster merger. Thus, this emission may constitute some form of `radio phoenix' \citep[see the review by][]{vanWeeren2019}. Radio phoenixes (and other similar revived fossil plasma phenomena) often exhibit an ultra-steep spectrum \citep[$\alpha \lesssim -2$; e.g.][]{deGasperin2017,Mandal2020,Ignesti2020b}, which is consistent with the upper limit to the spectral index derived from our ASKAP and ATCA data.  

One obvious candidate for the seed population of this ultra-steep spectrum source is the highly-extended head-tail radio galaxy directly west of the emission (source `RG3'). This galaxy lies at $z = 0.0629$, and is a member of the Western Core Component \citep[WCC; see][]{Dehghan2017}. The WCC is the more massive sub-cluster in the centre of Abell~3266, it is also relatively compact (compared to the Eastern Core Component) and has high velocity dispersion.

From Figure~\ref{fig:central}, the extended tail of this galaxy is about 30\% longer at 943~MHz than 2.1~GHz. Much of the faint tail is steep spectrum, with $\alpha_{\rm 943~MHz}^{\rm 2.1~GHz} \lesssim -1.5$. The extended tail blends seamlessly into the more extended diffuse central emission. Measured from our ASKAP image at 15~arcsec resolution (top left panel of Figure~\ref{fig:central}), the extended tail has a (projected) length of around 205~arcsec, equivalent to 238~kpc. From the lower-right panel of Figure~\ref{fig:central}, we see that the diffuse central emission is well-bounded by the edge of the low-entropy `spine' that runs through the cluster centre before deviating sharply to the NE. 

An alternative candidate for the origin of the diffuse synchrotron emission is the BCG itself, which lies at the boundary between the two central sub-clusters \citep{Dehghan2017}. The BCG has a known dumbbell morphology \citep{Henriksen2002,Finoguenov2006} that is clearly demonstrated in Figure~\ref{fig:central}. This morphology, combined with the high relative velocity of the BCG nucleus \citep[327~km~s$^{-1}$;][]{Finoguenov2006}, suggests that the BCG has been tidally disturbed during the cluster merger. We note the presence of extended optical emission in Figure~\ref{fig:central}, which broadly follows the morphology of the diffuse radio emission. Thus, the diffuse radio and optical emission may trace the same underlying phenomenon: material that has been disrupted by the close interaction of two galaxies which now constitute the dumbbell BCG and dispersed into the ambient medium.

\subsubsection{A mini-halo candidate?}
Historically, `mini-haloes' have typically been found almost exclusively in relaxed, cool-core clusters \citep[][]{Giacintucci2017,vanWeeren2019}. However, next-generation low-frequency observations have blurred the lines significantly, with new detections of multi-component mini-haloes hosted by moderately-disturbed clusters or clusters that show signs of significant core sloshing \citep{Venturi2017,Kale2019,Savini2019,Biava2021,Knowles2022,Riseley2022}.

The linear size of source D3, $240$~kpc, is typical of known mini-haloes, which are commonly around 100$-$300~kpc in size \citep{vanWeeren2019}, although next-generation observations are revealing that some mini-haloes are significantly larger, up to $\gtrsim0.5$~Mpc \citep{Savini2019,Biava2021,Riseley2022}. Mini-haloes frequently show an association with the radio BCG; from Figure~\ref{fig:central} (panels a and b), we see that several cluster-member radio galaxies, including the BCG, appear embedded in the diffuse emission.

With our measurement of the integrated flux density and our upper-limit spectral index of $\alpha_{\rm 943~MHz}^{\rm 2.1~GHz} = -2.54$, we use Equation~\ref{eq:radio_lum} to estimate an upper limit to the total radio power of $P_{\rm 1.4~GHz} = (2.04 \pm 0.32) \times 10^{22}$~W~Hz$^{-1}$. This extremely low radio power sits almost two orders of magnitude below established scaling relations \citep{Giacintucci2019,Savini2019}, although another similarly low-power mini-halo candidate ($P_{\rm 1.4~GHz} \sim 2 \times10^{22}$~W~Hz$^{-1}$) has recently been reported in the EMU Pilot Survey region by \cite{Norris2021}. 

It is entirely plausible that the sensitivity to ultra low surface brightness emission possessed by ASKAP is allowing us to explore new parameter space in the study of faint diffuse radio sources in clusters of galaxies. However, we note several counterpoints to the mini-halo interpretation. Firstly, our upper-limit spectral index is far steeper than is typically found for mini-haloes ($\alpha \sim -1$; e.g. \citealt{vanWeeren2019,Biava2021,Riseley2022}). 

Secondly, our \xmm{} temperature map, shown in panel (e) of Figure~\ref{fig:central} does not show strong evidence of a cool core in the region where the diffuse emission is found. \cite{Giacintucci2017} studied the occurrence rates for mini-haloes in a large sample of high-mass $(M \gtrsim 6\times10^{14}~\rm M_{\odot})$ galaxy clusters; the occurrence rate is high $(\sim80\%$) for cool-core clusters, and extremely low $(\sim0\%)$ for non-cool-core clusters. Only one non-cool-core cluster is known to host a mini-halo: Abell~1413 \citep{Govoni2009,Savini2019}. At X-ray wavelengths this cluster exhibits an elongated and slightly disturbed morphology \citep[e.g.][]{Pratt2002,Botteon2018b}, and exhibits evidence of larger-scale diffuse radio emission beyond the known mini-halo (Lusetti et al., in prep.; Riseley et al., in prep).

Abell~3266 has a high total cluster mass of $M = (6.64^{+0.11}_{-0.12})\times10^{14}$~M$_{\odot}$ \citep{Planck2016b} and shows no evidence of a cool core. As such, the presence of a mini-halo in this system is unlikely, but we cannot conclusively rule out that source D3 is a mini-halo.

\subsection{A low-brightness radio halo}
Previous searches for a large-scale radio halo in Abell~3266 have been inconclusive, largely due to contamination from discrete sources that could not be excised from lower-resolution data \citep{Bernardi2016,RiseleyThesis}. Recently, \cite{Duchesne2022} placed the tightest constraints on the presence of a radio halo in Abell~3266, deriving an upper limit of $P_{\rm 1.4~GHz} \lesssim 0.8 \times 10^{24}$~W~Hz$^{-1}$. Our new data provide the opportunity to perform a deeper search for a faint radio halo.

We applied the multi-resolution spatial filtering algorithm introduced by \cite{Rudnick2002b}\footnote{See also \cite{Rudnick2002a}.} to our 15~arcsec ASKAP mosaic, using a box size of 48~arcsec. This removes all of the flux from features in the image that are unresolved by our 15~arcsec beam. It removes progressively less of the flux as the scale sizes of the structures increase, reaching a shape-dependent average of $\sim$50\% of the flux on 48~arcsec scales. Since the filter resets the zero level on the map, we corrected the filtered image to an average flux density of zero in extended regions $\sim$15~arcmin to the NW and to the SE of the cluster centre. We then convolved the filtered image to a resolution of 48~arcsec.

\begin{figure*}
\begin{center}
\includegraphics[width=1.0\textwidth]{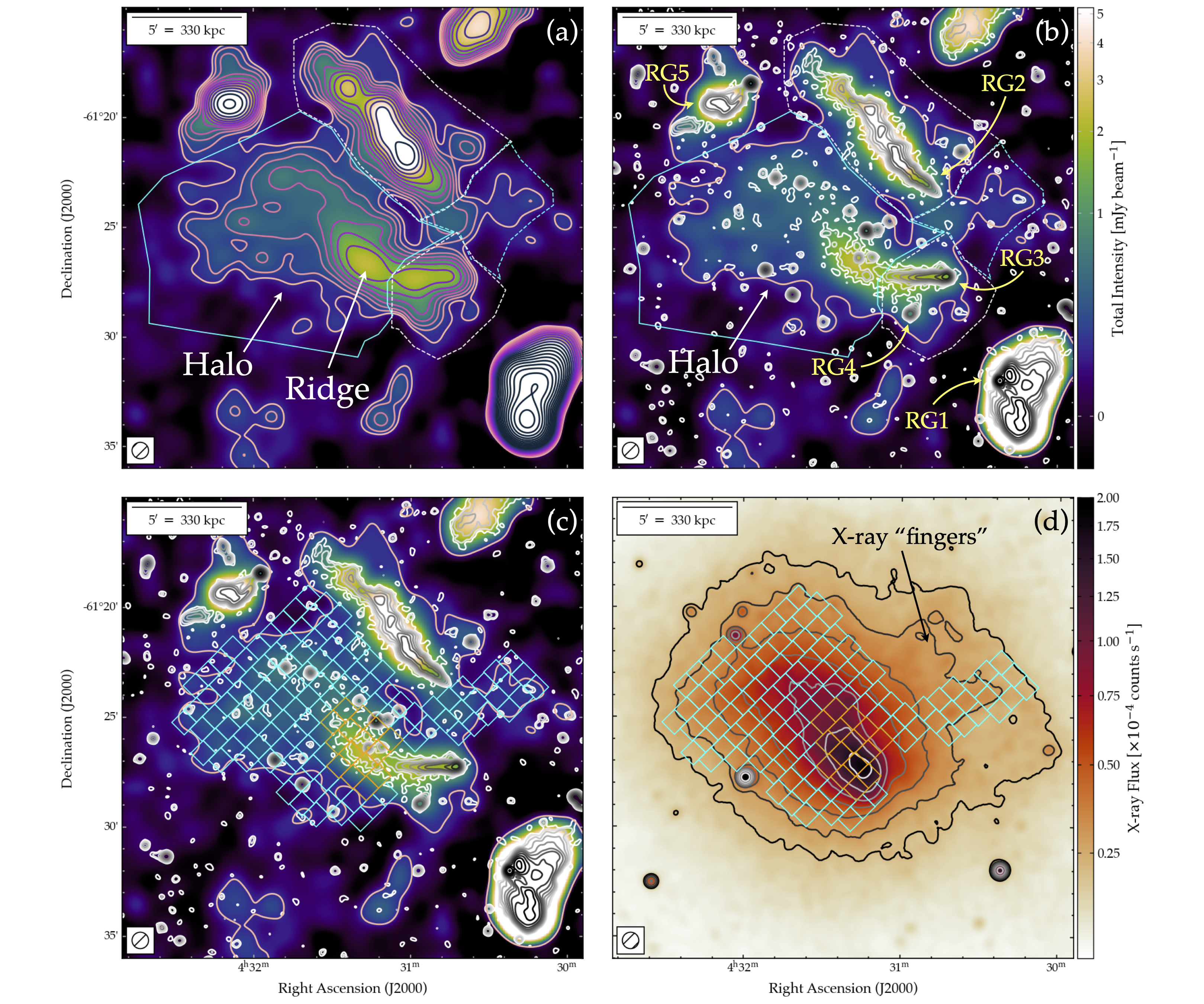}
\caption{The central region of Abell~3266. Panels (a), (b), and (c) show the our spatially-filtered 48~arcsec resolution ASKAP map at 943~MHz, using the colourbar adjacent to panel (b). Panel (a) shows the low-resolution contours starting at $3\sigma$ and scaling by $\sqrt{2}$, where $\sigma = 69.6~\upmu$Jy~beam$^{-1}$. Panels (b) and (c) show the $3\sigma$ low-resolution contour from panel (a) as well as our 15~arcsec ASKAP contours $3\sigma$ and scaling by factor $2$, for context. Panel (d) shows our \xmm{} mosaic image smoothed with a Gaussian of FWHM 48~arcsec, with representative contours starting at $0.2\times10^{-4}$~counts~s$^{-1}$ and scaling by $\sqrt{2}$. Regions in panels (a) and (b) show the area over which the integrated flux density measurements were derived. The boxes in panels (c) and (d) show the regions used to profile the radio/X-ray surface brightness correlation.}
\label{fig:filtered_map}
\end{center}
\end{figure*}

We show this low-resolution image of the central region of Abell~3266 in Figure~\ref{fig:filtered_map}, along with our \xmm{} X-ray surface brightness mosaic smoothed to the same resolution. Diffuse synchrotron emission fills much of the X-ray emitting volume of Abell~326, which we identify as a radio halo --- conclusively detected here for the first time. While much of this emission is associated with the ICM of Abell~3266, we also note residuals associated with the radio galaxies RG2 and RG3. The diffuse emission from the ICM fans out to the north, north-east, and east of the narrower band of diffuse emission reported at 15~arcsec resolution.

We also detect a region of diffuse emission associated with the `fingers' of X-ray emission reported by \cite{Henriksen2002}, extending to the north-west of the cluster centre. This X-ray surface brightness extension, highlighted in Figure~\ref{fig:filtered_map}, is also seen in the eROSITA images presented by \cite{Sanders2022}; the outer edge of the `fingers' appears as a sharp discontinuity in the edge-filtered X-ray maps in Figure~8 of \cite{Sanders2022}.

To determine the extent of the diffuse emission, we measured east to west across the region defined by the $3\sigma$ contour. Including the emission associated with the X-ray `fingers', the LAS is 16.5~arcmin (1.09~Mpc); conservatively excluding this emission, we measure a LAS of 10.5~arcmin (695~kpc). Both measurements are typical of the scale size of classical radio haloes.

\begin{table}
    \centering
    \caption{\label{tab:halo:flux} Integrated flux density of the diffuse radio halo. Note that here we quote only the statistical uncertainty; see text for discussion of systematic uncertainty.}
    \begin{tabular}{c c c c c}\hline
         Value & Size & $S_\text{943~MHz}$ & $S_\text{216~MHz}^{\alpha=-1.4}$ & $S_\text{216~MHz}^{\alpha = -1.2}$  \\
          & [kpc] & [mJy] & [mJy] & [mJy] \\\hline
         Minimum & 695 & $42.2 \pm 4.3 $ & $330 \pm 30$ & $250 \pm 30$ \\
         Maximum & 1090 & $67.8 \pm 6.2$ & $530 \pm 50$ & $400 \pm 40$ \\\hline
    \end{tabular}
\end{table}

\subsubsection{Flux density measurement}
The exact coverage of the diffuse radio halo is difficult to determine from the filtered ASKAP map due to confusion from extended components of RG2, RG3, and RG4. We measure the integrated flux density of the radio halo in two ways: (1) obtaining a minimum value assuming the emission region is bounded in the North by RG2 and in the West by RG3/4, and (2) obtaining a maximum value by assuming the diffuse emission extends towards the NW associated with the X-ray `fingers', and assuming the emission extends into the regions covering by RG2 and RG3/4. To estimate the flux density of the regions containing RG2 and RG3/4, we assume the surface brightness of those regions is equivalent to the mean surface brightness in the adjacent measured regions -- the `fingers' region for RG2 and the central region for RG3/4. 

The region over which measurement (1) (the `minimum' value) was derived is shown by the solid cyan polygon in panel (a) of Figure~\ref{fig:filtered_map}. Measurement (2) (the `maximum' value) also incorporates the dashed cyan polygon as well as the halo surface brightness as extrapolated across the dashed white polygons. Integrating above the $3\sigma$ contour for these regions we obtain $S_{\rm{943~MHz}}^{\rm{min}} = 42.2 \pm 4.3 \, (\pm 10.6)$~mJy and $S_{\rm{943~MHz}}^{\rm{max}} = 67.8 \pm 6.2 \, (\pm 17.0)$~mJy. 

When reporting the measurements, we quote two uncertainties: (i) standard statistical uncertainty and (ii) systematic uncertainty. This latter quantity is the dominant contributor to the error budget, and it arises from the uncertainty in determining the correct zero level, which comes from both the possibility of true faint emission very far from the cluster centre, and low-level background variations in the map. We estimate the uncertainty in the zero level to be $\sim$0.1, resulting in a 25 per cent uncertainty in the total flux.

Taking a typical radio halo spectral index of $\alpha = -1.4$ \citep{Duchesne2021c}, these 943~MHz measurements correspond to flux densities at 216~MHz of $330 \pm 30 \,(\pm 80)$~mJy and $530 \pm 50 \, (\pm 130)$~mJy, consistent with limits reported by \citet{Duchesne2022} for a radio halo of radius 500~kpc at the cluster centre. The larger, second set of uncertainties come from possible errors in the zero level. Measured and extrapolated flux densities are reported in Table~\ref{tab:halo:flux}.

Similarly, we substitute these minimum and maximum values, and the assumed spectral index $\alpha = -1.4$ in Equation~\ref{eq:radio_lum} to derive the minimum and maximum $k$-corrected 1.4~GHz luminosity. We find values of $P_{\rm 1.4~GHz}^{\rm min} = 1.94 \pm 0.20 (\pm 0.49) \times 10^{23}$~W~Hz$^{-1}$ and $P_{\rm 1.4~GHz}^{\rm max} = 3.11 \pm 0.29 (\pm 0.78) \times 10^{23}$~W~Hz$^{-1}$. 

Comparing these radio luminosity values against established scaling relations, we find that our measurements place the radio halo in Abell~3266 a factor $4.4\times$ to $7.1\times$ below the correlation plane for the $P_{\rm 1.4~GHz} / M_{500}$ relation derived by \cite{Duchesne2021c}, consistent with the upper limits from \cite{Duchesne2022}.

\subsubsection{Point-to-point correlation: the thermal/non-thermal connection}
As seen in Figure~\ref{fig:filtered_map}, the extent of the diffuse radio halo matches the extent of the ICM plasma (traced by the X-ray emission), implying a connection between the thermal and non-thermal components. Examining this connection via the correlation between radio surface brightness $(I_{\rm{R}})$ and X-ray surface brightness $(I_{\rm{X}})$, for example, can provide insights into the particle acceleration mechanism at work \citep{Govoni2001a}. This correlation takes the form ${\rm log} (I_{\rm{R}}) \propto b \, {\rm log} (I_{\rm{X}})$, where $b$ quantifies the slope. Physically, this slope relates to the relative distribution of non-thermal and thermal components, with a super-linear slope (i.e. $b > 1$) indicating that the magnetic field and/or CRe distribution is more peaked than the thermal plasma distribution, whereas a sub-linear slope (i.e. $b < 1$) indicates the converse.

\begin{figure}
\begin{center}
\includegraphics[width=0.95\linewidth]{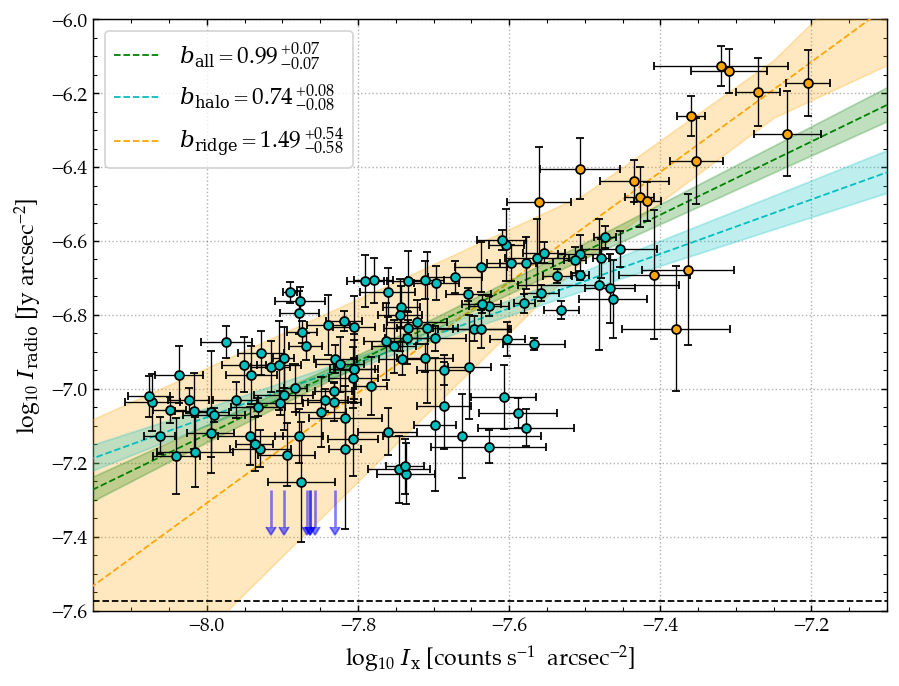}
\cprotect\caption{Radio/X-ray surface brightness correlation plane for the `radio halo' in Abell~3266. We report the radio surface brightness $(I_{\rm{R}})$ measured from our spatially-filtered 943~MHz ASKAP map at 48~arcsec resolution, and the X-ray surface brightness $(I_{\rm{X}})$ measured from our \xmm{} mosaic smoothed with a Gaussian of 48~arcsec FWHM. Markers are colourised according to extraction region (as per Figure~\ref{fig:filtered_map}). Upper limits ($2\sigma$) are shown by blue arrows. Dashed lines and shaded regions correspond to the best-fit trends from \textsc{linmix}, with the slopes shown in the legend, colourised according to population; the green line denotes the fit performed on \textit{all} regions together. The dashed horizontal line denotes the $1\sigma$ level from Figure~\ref{fig:filtered_map}. Error bars reflect the statistical errors on the measurements.}
\label{fig:halo_ptp}
\end{center}
\end{figure}

We profiled the $I_{\rm{R}}/I_{\rm{X}}$ connection by placing adjacent boxes of 48~arcsec width across the extent of the `radio halo'; these regions are also shown in Figure~\ref{fig:filtered_map}. We excluded any regions that clearly showed association with resolved radio galaxies in our 15~arcsec ASKAP map, to try to mitigate contamination from these sources. Figure~\ref{fig:halo_ptp} presents our point-to-point correlation, which appears to show a reasonably strong positive trend. To quantify the correlation, we fitted a linear relation (in log-log space) of the form:
\begin{equation}
    {\rm{log}}(I_{\rm{R}}) = a + b {\rm{log}}(I_{\rm{X}}),
\end{equation}

To determine the optimal values of $a$ and $b$, we used the MCMC implementation of \textsc{linmix} \footnote{\textsc{linmix} is available through \url{https://linmix.readthedocs.io/en/latest/src/linmix.html}} \citep{Kelly2007}. This software package performs Bayesian linear regression, accounting for uncertainties on both variables, intrinsic scatter and upper limits to the dependent variable (in this case $I_{\rm{R}}$). We take our `best-fit' values of $a$ and $b$ to be the median values of our MCMC chain, and use the 16th and 84th percentiles to determine the uncertainties. The correlation strength was quantified by measuring the Spearman and Pearson rank coefficients.

Fitting a single population to all our data, we find that the $I_{\rm{R}}/I_{\rm{X}}$ plane is described by a linear slope of $b_{\rm all} = 0.99 ^{+0.07}_{-0.07}$; this slope is shown by the dashed green line in Figure~\ref{fig:halo_ptp}, with the shaded region denoting the uncertainty. Overall, the data are strongly correlated, as we find Spearman (Pearson) coefficients of $r_S = 0.71$ $(r_P = 0.76)$. However, a number of regions toward higher X-ray surface brightness show clear departure from a single relation; as such, we also searched for substructure by subdividing our data. We performed our \textsc{linmix} regression on the `ridge' --- the emission coincident with source D3 detected at 15~arcsec (orange regions) --- and the larger-scale `halo' --- those regions which only appear at lower resolution after applying the spatial filter (cyan regions) --- separately.

\begin{figure*}
\begin{center}
\includegraphics[width=0.99\textwidth]{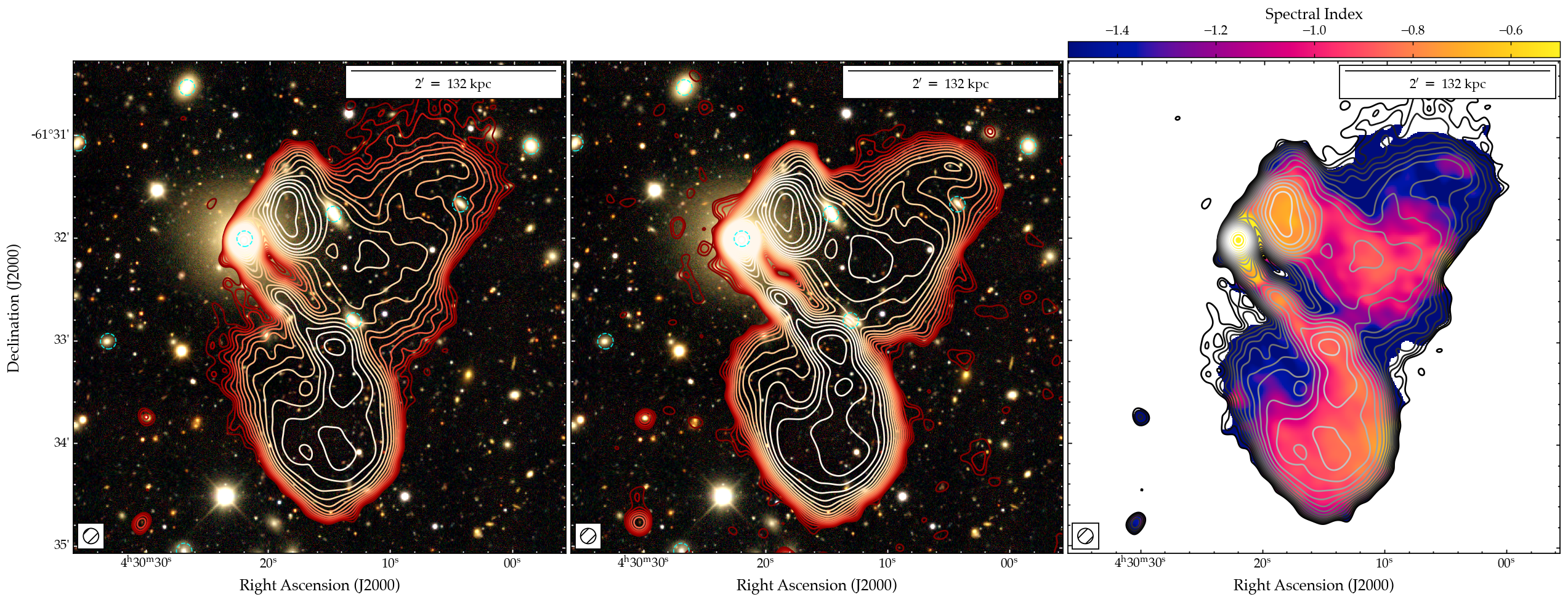}
\caption{Images of the wide-angle tail radio galaxy to the SE of Abell~3266 (source `RG1' in Figure~\ref{fig:A3266_RGB} at a resolution of 9~arcsec. \emph{Left} and \emph{centre} panels show our RGB image constructed from DES, with 943~MHz ASKAP and 2.1~GHz ATCA contours overlaid respectively. \emph{Right} panel shows the radio spectral index, with ASKAP contours overlaid. All contours start at $4\sigma$ and scale by a factor of $\sqrt{2}$. Known cluster-member galaxies are identified by dashed cyan circles.}
\label{fig:WAT}
\end{center}
\end{figure*}

We found that the `ridge' regions are described by a super-linear slope of $b_{\rm ridge} = 1.49 ^{+0.55}_{-0.58}$, whereas the `halo' regions are described by a sub-linear slope of $b_{\rm halo} = 0.72 ^{+0.09}_{-0.08}$. Each of these subsets shows a strong/moderate-to-strong correlation: for the `halo', $r_S = r_P = 0.61$, and for the `ridge', $r_S = 0.66$ and $r_P = 0.50$. While the uncertainty on $b_{\rm ridge}$ is significant, and thus the slope is not inconsistent with a linear trend, the values of $b_{\rm ridge}$ and $b_{\rm halo}$ are significantly different. As such, these boxes may trace regions that have a different connection between the thermal and non-thermal components of the ICM.

\subsubsection{On the nature of the radio halo in Abell~3266}
In general, classical radio haloes have been found to observe \emph{sub-linear} scaling relation, i.e. $b < 1$ \citep{Govoni2001a,Govoni2001b,Giacintucci2005,Hoang2019,Hoang2021,Botteon2020b,Rajpurohit2021b,Rajpurohit2021c,Duchesne2021c}, whereas mini-haloes have typically been found to follow a \emph{super-linear} scaling relation, i.e. $b > 1$ \citep{Govoni2009,Ignesti2020,Ignesti2022,Biava2021,Riseley2022}. However, there are contrary cases: \cite{Timmerman2021} reported that the mini-halo in the Perseus cluster shows sub-linear scaling, whereas \cite{deGasperin2022} reported that the classical halo in Abell~3667 shows super-linear scaling in the radio/X-ray point-to-point correlation.

The linear slope of $b = 0.99 ^{+0.07}_{-0.07}$ from our single-population analysis is consistent with the typical $I_{\rm{R}}/I_{\rm{X}}$ plane behaviour for radio haloes. However, our two-population analysis shows tentative evidence of substructure: the smaller-scale emission traced by the orange regions follows a super-linear trend, typical of mini-haloes, whereas the larger-scale emission traced by the cyan regions follows a clearly sub-linear trend, typical of haloes. 

Multi-component radio haloes are uncommon but not unprecedented in the literature. The complex merging cluster Abell~2744 also hosts a multi-component radio halo, where the different components follow distinctly different trends in the $I_{\rm{R}}/I_{\rm{X}}$ \citep{Rajpurohit2021c}. Another example is the cluster Abell~2142, which is dynamically disturbed due to merger activity but still hosts a cool core \citep{Wang2018}. Abell~2142 also hosts a multi-component radio halo, which has alternatively been classified as a `mini-halo-plus-halo' structure \citep{Venturi2017}. Several other clusters have recently been detected which host both a mini-halo and larger-scale diffuse radio emission on scales up to $\gtrsim0.5$~Mpc \citep[][Lusetti et al., in prep.; Riseley et al., in prep.]{Savini2018,Savini2019,Raja2020,Biava2021,Riseley2022}. Some of these are known cool-core clusters, some host no clear cool core; all exhibit some level of large-scale disturbance which could inject large-scale turbulence either via core sloshing or minor merger activity. 

For example, the cool-core cluster RX~J1720.1$+$2638 hosts both a mini-halo and larger-scale diffuse emission underlying much of the cluster volume \citep{Biava2021}. In this system, the mini-halo exhibits a clearly \textit{super}-linear correlation in the $I_{\rm{R}}/I_{\rm{X}}$ plane, whereas the larger-scale diffuse emission exhibits a demonstrably \textit{sub}-linear correlation; see \cite{Biava2021} for details. This behaviour is consistent with the evidence seen in Figure~\ref{fig:halo_ptp}.

Overall, our findings for Abell~3266 are consistent with the interpretation that we are viewing multi-component radio halo, although the nature of the smaller-scale component remains unclear at present. Given the highly complex merger event occurring in Abell~3266, further work is highly motivated: deeper observations are required to explore the point-to-point correlation in more detail, whereas deeper theoretical exploration of the effect of complex merger events on cluster-scale turbulence and the thermal/non-thermal connection would be required to aid in interpreting our results.

\section{Resolved radio galaxies}\label{sec:tailed_galaxies}
Abell 3266 hosts a number of other spectacular bent-tail radio galaxies, which are resolved in exquisite detail in our new ASKAP and ATCA maps. These will be discussed in the following subsections.

\subsection{RG1: the wide-angle tail radio galaxy}
Source RG1 is perhaps the most spectacular radio galaxy hosted by Abell~3266. It is a wide-angle tail radio galaxy, identified variously in the literature as MRC~0429$-$616 or PKS~0429$-$61, for example, hosted by the galaxy J043022.01$-$613200.2 at redshift $z = 0.0557$ \citep{Dehghan2017}. From the optical substructure analysis presented by \citeauthor{Dehghan2017}, this galaxy is a member of the WCC and lies toward the outermost edge of this sub-cluster. We present our images of this source at 9~arcsec resolution in Figure~\ref{fig:WAT}.

\subsubsection{Radio Continuum Morphology}
RG1 has a compact, bright radio core with jets that point NW and SW with a projected opening angle of around 95~degree. The NW jet terminates some 20~arcsec (21~kpc) beyond the core, expanding into a relatively bright clump of diffuse radio emission that fades into the larger-scale western lobe. Conversely, the SW jet extends much further, with two possible hotspots around 38~arcsec (39.5~kpc) and 77~arcsec (80~kpc) from the core. The outer hotspot marks roughly where the jet expands into the more diffuse southern lobe.

Our high-resolution radio maps presented in Figure~\ref{fig:WAT} clearly demonstrate that this cluster-member galaxy has undergone a dynamic history and/or experienced complex interactions with the cluster environment. The radio morphology of the lobes is complex: the southern lobe extends to the SW before bending clockwise and sweeping back upon itself, whereas the western lobe extends in a SW direction before turning to the NW. We also note a clear `hook'-like feature toward the outer edge of the western lobe that is detected by both ASKAP and the ATCA. Finally, we do not clearly detect the western fossil detected by \cite{Duchesne2022} beyond the westernmost edge of the western lobe. This feature is barely visible by-eye in the ASKAP image at 15~arcsec (Figure~\ref{fig:A3266_fullres}) but it is not significant compared to the local noise. The non-detection in our ASKAP and ATCA maps is consistent with the ultra-steep spectrum measured by \citeauthor{Duchesne2022} for this fossil source ($\alpha = -2.4 \pm 0.4$).

\subsubsection{Spectral index distribution}
The right panel of Figure~\ref{fig:WAT} presents the spectral index map of RG1 between 943~MHz and 2.1~GHz. Due to the high signal-to-noise achieved by both our ATCA and ASKAP maps, the uncertainty in the spectral index map for RG1 is highly uniform, with values between $0.08$ and $0.09$; as such, we do not present a map of the spectral index uncertainty. 

From the spectral index profile, the core shows a very flat spectrum ($\alpha_{\rm 943~MHz}^{\rm 2.1~GHz} = -0.33$); the jets exhibit fairly typical synchrotron spectra, with $\alpha_{\rm 943~MHz}^{\rm 2.1~GHz} = -0.6$ to $-0.8$. The jet that feeds the southern lobe shows variations along its length, with flatter-spectrum regions ($\alpha_{\rm 943~MHz}^{\rm 2.1~GHz} \simeq -0.85$) that align well with the hotspots in total intensity and steeper-spectrum regions ($\alpha_{\rm 943~MHz}^{\rm 2.1~GHz} \sim -0.95$) between the hotspots. This behaviour is typical of active jets. The southern lobe shows a spectral index gradient that broadly follows the surface brightness gradient (with some fluctuations visible). 

For the western lobe, the bright `clump' (where the NW jet presumably terminates) shows a typical synchrotron spectral index $\alpha_{\rm 943~MHz}^{\rm 2.1~GHz} = -0.65$ to $-0.85$, similar to the southern jet. Beyond the `clump', the spectral index gradient broadly follows the radio surface brightness gradient: the spectral index varies from around $\alpha_{\rm 943~MHz}^{\rm 2.1~GHz} \sim -0.85$ at the edge of the clump to $-1.4 \lesssim \alpha_{\rm 943~MHz}^{\rm 2.1~GHz} \lesssim -1.0$ toward the edge of the lobe. The `hook' feature identified in the radio surface brightness also shows a flatter spectral index of $\alpha_{\rm 943~MHz}^{\rm 2.1~GHz} \sim -1.3$ compared to the lower surface brightness regions at the extremity of the southern lobe, which show an ultra-steep spectral index (typically $\alpha_{\rm 943~MHz}^{\rm 2.1~GHz} \simeq -1.5$ to $-2$).

Comparing our spectral index map with the MeerKAT in-band spectral index map presented in Figure~1.B of \cite{Knowles2022}, we observe similar spectral behaviour on the whole. However, we also note a number of differences. This is primarily visible in the centroid of the southern lobe, where we measure a typical spectral index of $\alpha_{\rm 943~MHz}^{\rm 2.1~GHz} = -0.88$ (at 9~arcsec resolution) whereas the in-band MeerKAT spectral index reaches values as steep as $\alpha_{\rm 0.9~GHz}^{\rm 1.67~GHz} = -2.02$ (at $7.1 \times 6.7$~arcsec resolution). Similar differences are seen at the outer edge of the bright `clump' in the western lobe. This discrepancy is too significant to be attributable to measurement uncertainties alone. Given our use of \textsc{ddfacet}'s \textsc{ssd} algorithm, which jointly deconvolves emission on a variety of scales and therefore likely recovers the complex spatial distribution of the extended emission from this radio galaxy to a better degree, we suggest that the behaviour seen in the MeerKAT in-band spectral index map is unphysical, and our spectral index map provides a more accurate description of the underlying profile.

\subsubsection{On the nature of RG1}
Our high-resolution images of RG1 have provided unprecedented detail into the morphology and spectral properties of this active radio galaxy. However, they also raise several questions. What is the nature of the bright `clump' in the western lobe, and does it correspond to the termination point of the NW jet where it decelerates and expands into the western lobe? What is the dynamical history of this radio galaxy? 

\cite{DehghanThesis} attempted to understand the nature of RG1, suggesting that the overall motion of the host galaxy reversed some $\sim80$~Myr ago. However, numerical simulations of AGN in complex merging clusters would be required to understand more from the theoretical perspective. In order to explore the evolutionary history of this source in detail, we would require additional data at both higher and lower frequencies that is capable of achieving $\sim10$~arcsec resolution or better. MeerKAT observations with the UHF-band or S-band receiver systems, and/or higher-frequency data from the ATCA 4cm band would be suitable, but lower-frequency data at sufficient resolution will not be available until the advent of SKA-Low.

\begin{figure}
\begin{center}
\includegraphics[width=0.99\linewidth]{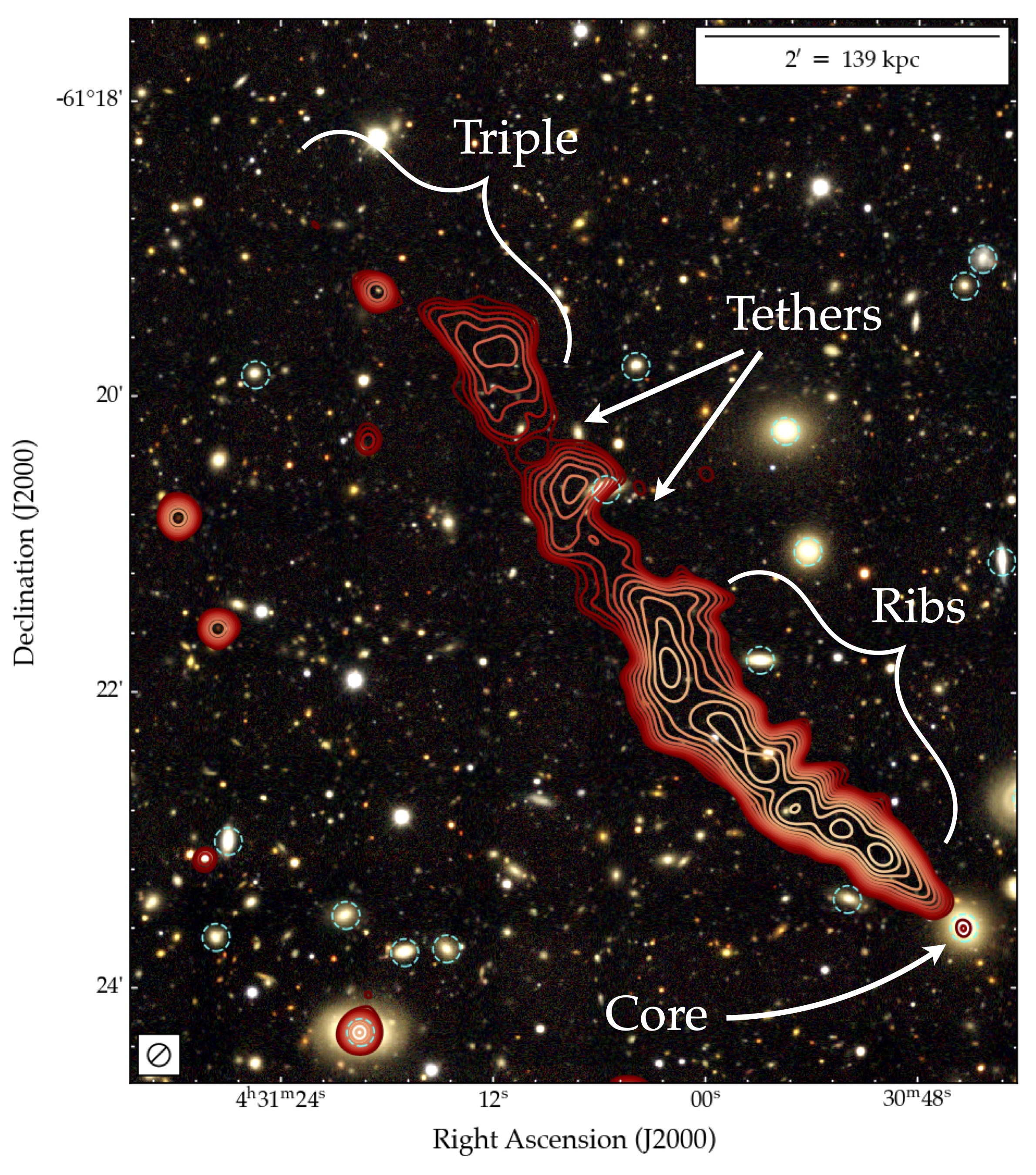}
\cprotect\caption{Image of source `RG2', the complex radio galaxy to the NW of Abell~3266 \citep[dubbed \emph{`MysTail'} by][]{Rudnick2021}, likely associated with the galaxy J043045.37$-$612335.8 at $z = 0.0628$ \citep{Dehghan2017}. Colormap shows our RGB image constructed from DES, with cluster-member galaxies highlighted by dashed cyan circles. Contours show the radio surface brightness measured by the ATCA at 2.1~GHz and 9~arcsec resolution, starting at $4\sigma$ and scaling by a factor $\sqrt{2}$. Labels identify key features which are referred to in the text. The north-eastern component of the triple is undetected by the ATCA, suggesting a highly steep spectral index.}
\label{fig:MysTail_zoom}
\end{center}
\end{figure}

\subsection{RG2: a complex radio galaxy}
\begin{figure*}
\begin{center}
\includegraphics[width=0.99\textwidth]{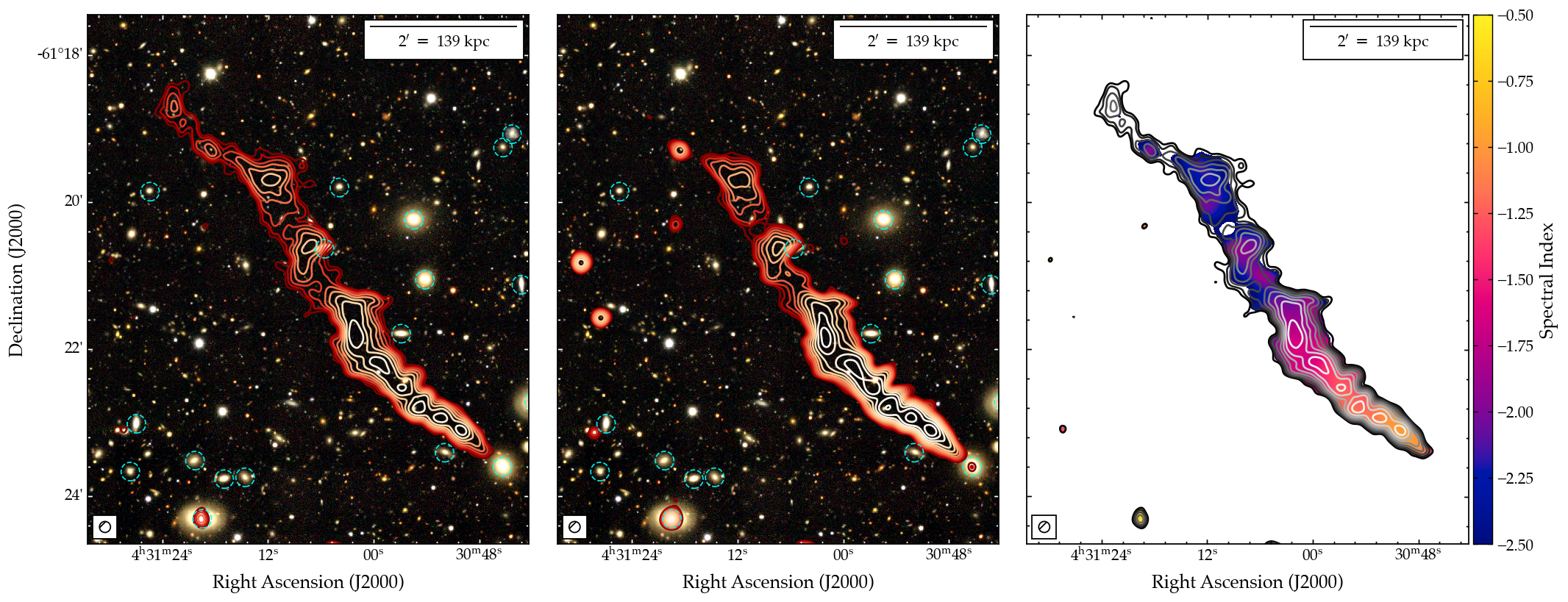}
\includegraphics[width=0.99\textwidth]{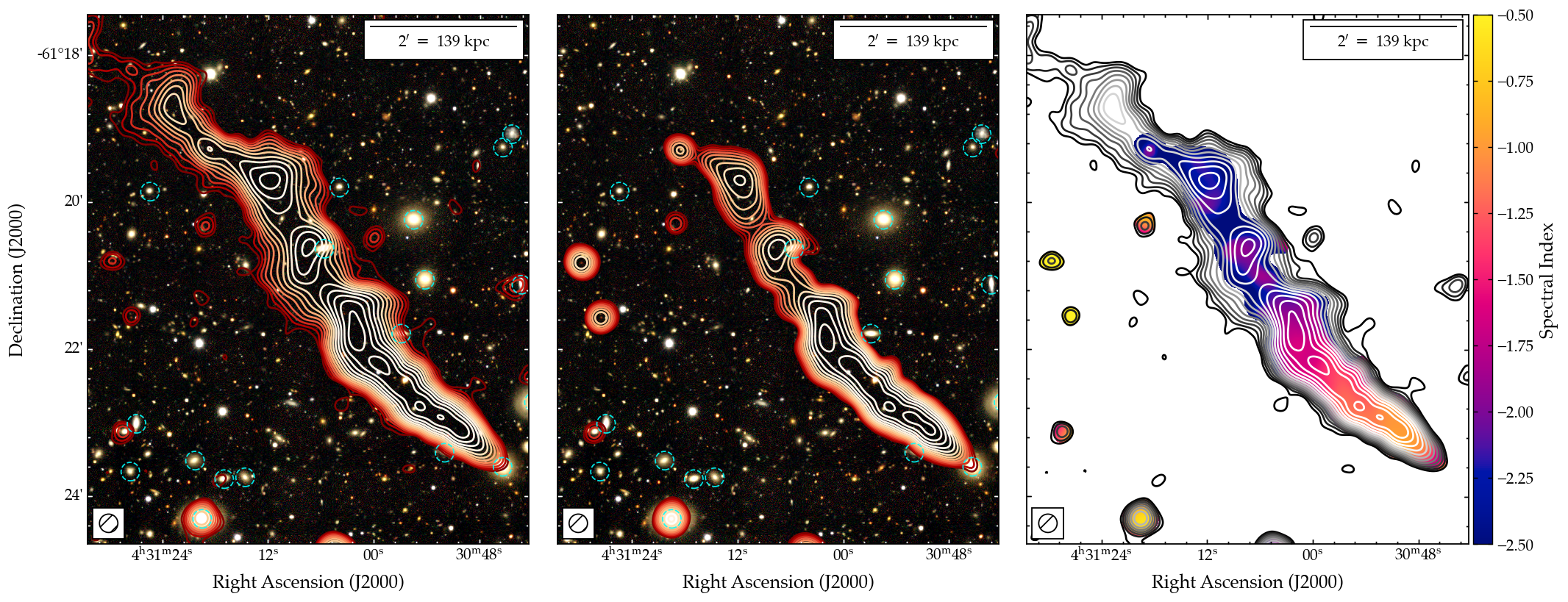}
\cprotect\caption{Images of the complex radio galaxy to the NW of Abell 3266 (source `RG2' in Figure~\ref{fig:A3266_RGB}). \emph{Top row} shows images at 9~arcsec resolution, \emph{bottom row} shows images at 15~arcsec resolution. \emph{Left} and \emph{centre} panels show RGB images from DES with 943~MHz ASKAP and 2.1~GHz ATCA contours overlaid respectively. \emph{Right} panel shows the radio spectral index, with ASKAP contours overlaid. All contours start at $4\sigma$ and scale by a factor of $\sqrt{2}$. }
\label{fig:MysTail_Overlay}
\end{center}
\end{figure*}

Source RG2 lies to the NW of the cluster centre. This source was first identified in relatively shallow narrow-band observations \citep{1999Murphy,1999PhDTReid,Murphy2002}. However, the nature of this source has historically remained unclear due to the poor sensitivity and resolution of previous observations. It has previously been hypothesised that this source represents a dying radio galaxy tail, possibly associated with the elliptical galaxy to the immediate SW \citep{1999Murphy} or alternatively a radio relic \citep{DehghanThesis}. Most recently, \cite{Rudnick2021} present a detailed study of this source (dubbed `\emph{MysTail}' by \citeauthor{Rudnick2021}) using data from the MeerKAT GCLS \citep{Knowles2022}. The MeerKAT images presented by \cite{Rudnick2021} and \cite{Knowles2022} clearly demonstrate that this source is a radio galaxy --- albeit one that exhibits a highly complex morphology --- or may, in fact, comprise multiple radio galaxies. Figure~\ref{fig:MysTail_zoom} presents a zoom on our ATCA map of this source at 9~arcsec resolution, with the main features of interest highlighted for reference. We also present ASKAP and ATCA maps of this source at both 9~arcsec and 15~arcsec resolution, along with the spectral index map, in Figure~\ref{fig:MysTail_Overlay}. 

In our 9~arcsec resolution ATCA map, we detect a compact radio counterpart to the cluster-member galaxy J043045.37$-$612335.8 \citep[$z = 0.0628$;][]{Dehghan2017}, at the south-west tip of this source. This is the candidate optical host originally proposed by \cite{1999Murphy}, although \citeauthor{Dehghan2017} provide the first spectroscopic redshift for this source. The compact radio source is extremely faint, with an (unresolved) flux density of $S_{\rm 2.1~GHz} = 90 \pm 21~\upmu$Jy, and it is undetected in our 9~arcsec resolution ASKAP image, suggesting a flat or inverted spectral index. 

\cite{Knowles2022} report a peak flux density measurement of around $65~\upmu$Jy beam$^{-1}$ in the full-resolution ($\sim8$~arcsec) MeerKAT GCLS image of Abell~3266. Taking this to be the flux density of the (unresolved) radio core, this implies an inverted spectral index around $\alpha_{\rm 943~MHz}^{\rm 1.28~GHz} \sim +0.7$. We detect no significant emission corresponding to jets from this proposed radio core that feed into the larger-scale ribbed tail seen in the images presented here as well as those presented by \citeauthor{Rudnick2021} and \citeauthor{Knowles2022}.

\subsubsection{Integrated spectrum and radio luminosity}
\begin{figure}
\begin{center}
\includegraphics[width=0.95\linewidth]{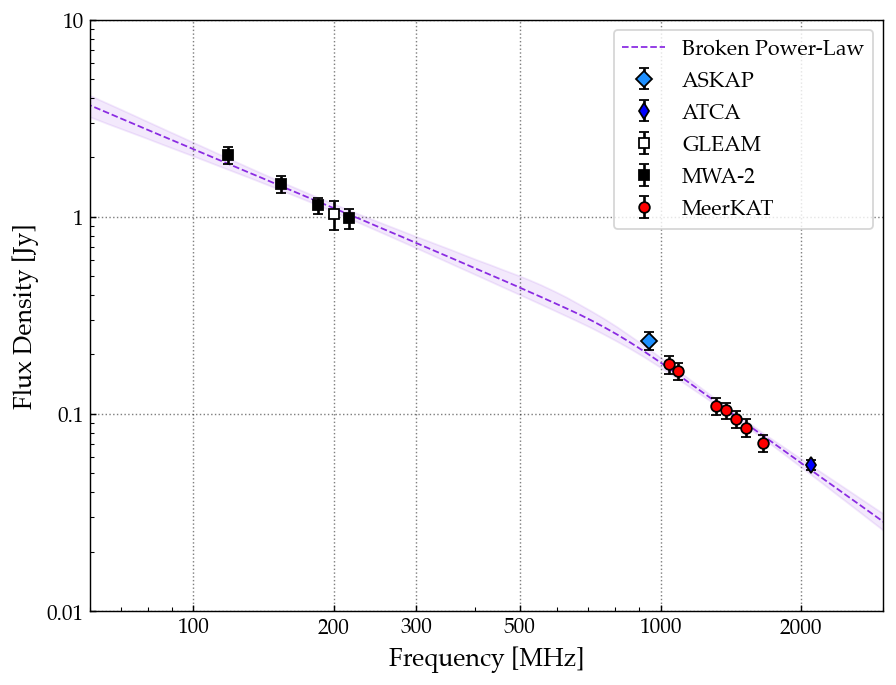}
\cprotect\caption{Spectral energy distribution (SED) for the complex tailed radio galaxy to the north-west of Abell 3266 (Source RG2; shown in Figure~\ref{fig:Fossil_Overlay}). Our new ASKAP and ATCA measurements are shown as blue diamonds, in-band spectral measurements from the MeerKAT GCLS data are shown in red, and datapoints from the MWA from \cite{Duchesne2022} are shown as squares. Our best-fit broken power-law spectrum is shown by the dashed violet line; the shaded region denotes the $1\sigma$ uncertainty.}
\label{fig:MysTail_SED}
\end{center}
\end{figure}

To measure the total integrated flux of RG2, we defined a region corresponding to the $4\sigma$ contour of our ASKAP map at 15~arcsec resolution. Integrating over this area, we measure a total flux of $S_{\rm int, 943~MHz} = 235 \pm 24$~mJy with ASKAP and $S_{\rm int, 2.1~GHz} = 55 \pm 3$~mJy with the ATCA. This implies an integrated spectral index $\alpha_{\rm 943~MHz}^{\rm 2.1~GHz} = -1.81 \pm 0.14$. We combine our new measurements with the lower-frequency data presented by \cite{Duchesne2022} and measurements made on the public GCLS images, to investigate the broad-band integrated spectral index of RG2; this is presented in Figure~\ref{fig:MysTail_SED}. Overall, the SED exhibits evidence of departure from a single power-law behaviour, so we attempted to model the behaviour with \textsc{synchrofit} using physically-motivated models. We modelled the behaviour using the `CI-on' and `CI-off' models, which would represent a radio galaxy observed in an active or remnant phase, respectively. Neither model provided a good description of our data, so we instead defer to a simple broken power-law fit.

This broken power-law fit is also presented in Figure~\ref{fig:MysTail_SED}. Our modelling yields a low-frequency spectral index of $\alpha_{\rm low} = -1.01^{+0.13}_{-0.07}$ below the break frequency and $\alpha_{\rm high} = -1.67^{+0.13}_{-0.12}$. The break frequency itself is $\nu_{\rm c} = 0.77^{+0.16}_{-0.17}$~GHz. While the break frequency is relatively poorly constrained, and would require observations with the MeerKAT UHF-band receiver system to constrain more accurately, $\alpha_{\rm low}$ and $\alpha_{\rm high}$ are inconsistent, suggesting that the observed spectral steepening is real. Given our cosmology, we use Equation~\ref{eq:radio_lum} and our best-fit model to derive the radio power, yielding $P_{\rm 1.4~GHz} = (8.05 \pm 0.02) \times 10^{23}$~W~Hz$^{-1}$.

\subsubsection{Core prominence}
We investigated the core prominence (CP) for RG2, to determine whether it is typical of active or remnant radio galaxies. Using our 2.1~GHz measurements of the integrated flux density and the core flux density, we derive a value of $\rm CP_{2.1~GHz} = 1.6\times10^{-3}$. This is entirely consistent with known samples of remnant radio galaxies \citep{Jurlin2021,Quici2021}. Given the lack of clear jets at 2.1~GHz, as well as the inverted spectral index for the core and the spectral steepening observed in the SED, the question arises: could this represent a phase of restarted AGN activity?

\subsubsection{The ribs and tethers}
From our 9~arcsec resolution images presented in Figures~\ref{fig:MysTail_zoom} and \ref{fig:MysTail_Overlay}, we confirm the `ribbed tail' morphology reported by \cite{Knowles2022} (their Figure~19). At this resolution, we find that the `ribs' are uniformly around 20~arcsec to 30~arcsec wide at half-intensity at both 943~MHz and 2.1~GHz, similar to the width seen by MeerKAT \citep{Rudnick2021,Knowles2022}. 

Examination of the spectral index map presented in the top-right panel of Figure~\ref{fig:MysTail_Overlay} reveals substructure in the spectral index profile of the ribbed tail. There is an overall gradient following the `spine' of the tail from around $\alpha_{\rm 943~MHz}^{\rm 2.1~GHz} \simeq -0.8$ nearest to the proposed optical host to $\alpha_{\rm 943~MHz}^{\rm 2.1~GHz} \simeq -1.8$ at the north-eastern end where the ribbed tail fades into the region of the `tethers'. This is slightly steeper than the the behaviour reported by \cite{Rudnick2021}, who measure a spectral index of $\alpha_{\rm 0.9~GHz}^{\rm 1.67~GHz} = -0.6$ near the core and $\alpha_{\rm 0.9~GHz}^{\rm 1.67~GHz} \sim -1.4$ where the tail bends.

These `tethers' reported by \cite{Rudnick2021} and \cite{Knowles2022} are also recovered by both ASKAP and the ATCA at 9~arcsec resolution. At 15~arcsec resolution, these blend into each other and form a more contiguous bridge between the brighter knots of the extended RG2 tail. Where the surface brightness is recovered at sufficient signal-to-noise ratio, we measure an ultra-steep spectral index of $\alpha_{\rm 943~MHz}^{\rm 2.1~GHz} \sim -2.0 \pm 0.2$ for the brighter `fingers' of emission with an even steeper spectral index of $\alpha_{\rm 943~MHz}^{\rm 2.1~GHz} \lesssim -2.5$. Again, our measurements are consistent with the in-band measurements of $\alpha_{\rm 0.9~GHz}^{\rm 1.67~GHz} \sim -2$ to $-3.4$ reported by \citeauthor{Rudnick2021}

\subsubsection{The triple}
One of the open questions posed by \cite{Rudnick2021} and \cite{Knowles2022} regards the nature of the `triple' at the north-eastern end of RG2. Two scenarios exist: the simplest is that the diffuse emission of the `triple' is simply a continuation of the diffuse emission of the ribbed tail associated with the cluster-member radio galaxy J043045.37$-$612335.8. 

As seen in the lower panels of Figure~\ref{fig:MysTail_Overlay}, ASKAP recovers significant diffuse emission associated with all components of the triple, with a narrow tail extending further to the north-east. At lower frequencies, the MWA images presented by \cite{Duchesne2022} show that this diffuse tail extends yet further to the north-east. The spectral index in this outermost region must be steep, as it is largely undetected by ASKAP at 943~MHz.

However, the presence of the compact radio component visible in our 2.1~GHz ATCA map would appear to be an astonishing coincidence. This compact radio component has a flux density $S_{\rm 2.1~GHz} = 0.48 \pm 0.04$~mJy, and is coincident with a galaxy (DES~J043118.45$-$611917.9) at a photometric redshift $z_{\rm phot} = 0.722 \pm 0.129$ \citep{DES-photoz}; if this is the core of the `triple', then it would constitute a giant radio galaxy spanning a projected linear size of 1.03~Mpc \citep[as posited by][]{Rudnick2021}. The compact radio component is difficult to separate from the diffuse tail seen by ASKAP at 943~MHz, though a relatively compact component is seen by ASKAP at 9~arcsec resolution; at 15~arcsec resolution, diffuse emission dominates in this region.

At 9~arcsec resolution, we measure a spectral index of $\alpha = -1.18 \pm 0.14$ between 943~MHz and 2.1~GHz for the proposed core of the triple, from the map presented in the upper-right panel of Figure~\ref{fig:MysTail_Overlay}. However, this may be biased toward a steep spectrum by the diffuse emission overlaying this location. \cite{Rudnick2021} measure a peak flux density of 0.64~mJy~beam$^{-1}$ for the core, which suggests a significantly flatter core spectral index of around $\alpha \simeq -0.58$ between MeerKAT at 1.28~GHz and the ATCA at 2.1~GHz.

\subsubsection{On the nature of RG2}
The overall extent of RG2, measured from our 15~arcsec ASKAP image (lower-left panel of Figure~\ref{fig:MysTail_Overlay}) is 8.7~arcmin or 608~kpc at $z = 0.0628$. Like many other tailed radio galaxies that reside in cluster environments, RG2 exhibits a complex morphology with common features like ribs and tethers \citep[e.g.][]{Wilber2018,Wilber2019,Rajpurohit2022,Ignesti2022}. The latter of these examples, the head-tail galaxy GB6~B0335$+$096 is hosted by the galaxy cluster 2A~0335$+$096. LOFAR observations of this cluster presented by \cite{Ignesti2022} have revealed that the Mpc-scale tail of GB6~B0335+096 shows clumpy knotted regions along its length connected by regions of fainter diffuse emission. These `knots' show a flatter spectral index, and these regions likely trace instabilities between the tail and the ambient medium, being gently re-accelerated. 

\begin{figure*}
\begin{center}
\includegraphics[width=0.99\textwidth]{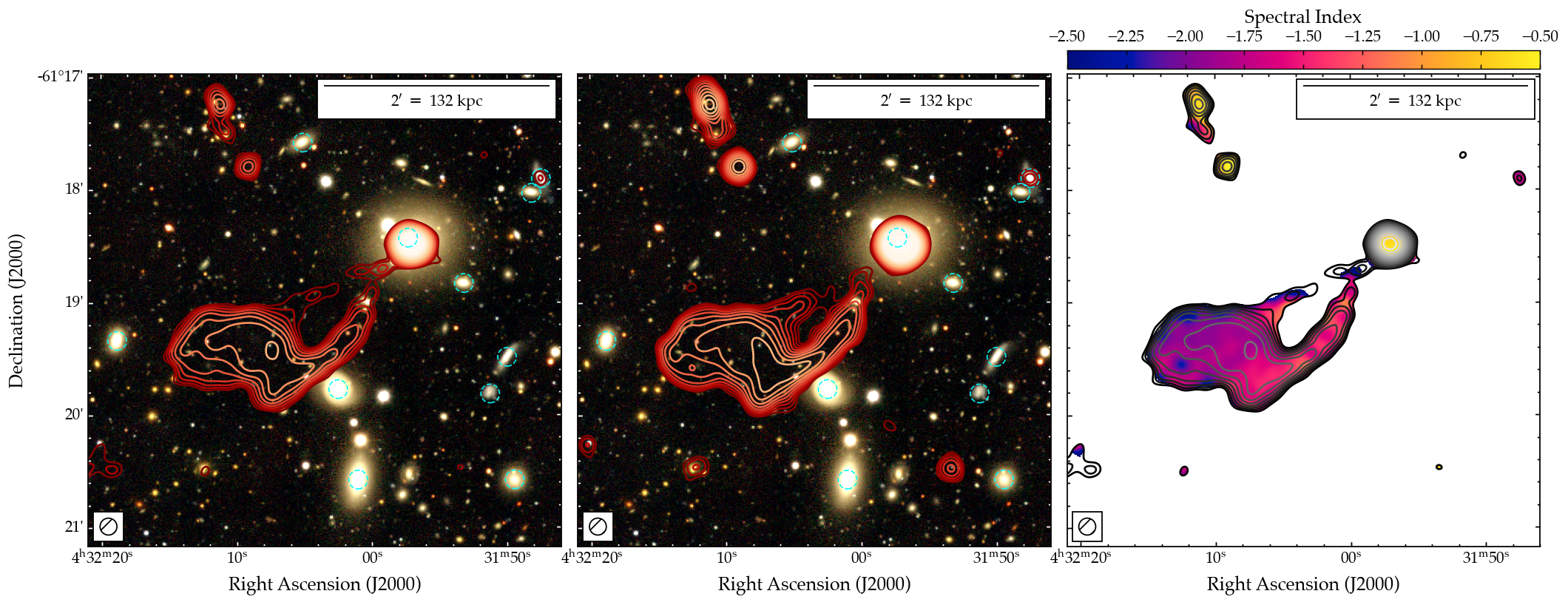}
\cprotect\caption{Images of the tailed radio galaxy to the NE of Abell 3266 (source `RG5' in Figure~\ref{fig:A3266_RGB}) at 9~arcsec resolution. The \emph{left} and \emph{centre} panels show RGB images from DES with 943~MHz ASKAP and 2.1~GHz ATCA contours overlaid respectively. \emph{Right} panel shows the radio spectral index, with ASKAP contours overlaid. All contours start at $4\sigma$ and scale by a factor of $\sqrt{2}$. }
\label{fig:RG5_Overlay}
\end{center}
\end{figure*}

Overall, the similarities between the morphological and spectral properties shown by RG2 and these similar sources strongly support the interpretation that this is a single contiguous radio galaxy associated with the cluster-member galaxy J043045.37$-$612335.8 at $z = 0.0628$. The presence of the compact radio source `interloper' that appears to be associated with DES~J043118.45$-$611917.9 is curious and motivates spectroscopic follow-up, but given the available multi-wavelength data we concur with the interpretation favoured by \cite{Rudnick2021}, that RG2 forms a single contiguous radio galaxy, with the complex morphology --- the ribs and tethers --- being generated by motions in the ICM. However, the origin of the various unusual features described by \cite{Rudnick2021} are not currently understood, and further work (both observational and theoretical) is motivated.

\subsection{RG5: an inclined tail?}
To the north-east of Abell~3266 lies the source identified in Figure~\ref{fig:A3266_RGB} as RG5. We present a zoom on this object in Figure~\ref{fig:RG5_Overlay}, which shows our ASKAP and ATCA images at 9~arcsec resolution, along with the spectral index map.

\subsubsection{Radio continuum properties}
RG5 is another somewhat complex radio galaxy. The core appears to be associated with the cluster-member radio galaxy J043157.29$-$611825.8 at $z = 0.059$, which is a member of Structure 2 \citep[][]{Dehghan2017}. Based on the optical analysis performed by \citeauthor{Dehghan2017}, Structure 2 is a radial filamentary structure with broad velocity dispersion, and has two concentrations of galaxies that are separated in redshift, suggesting a dynamically active environment.

The radio morphology of the extended jets and tails appears to support this interpretation. As is visible in Figure~\ref{fig:RG5_Overlay}, we detect what appear to be two jets extending from the compact (and bright) core of RG5. Both jets extend to the south-east, suggesting that this source belongs to the general `bent-tail' class although projection effects mean we cannot determine whether this may be a head-tail, a narrow-angle tail or a wide-angle tail radio galaxy.

The jets exhibit a striking asymmetry in surface brightness: the southern jet has on average a factor $\sim4\times$ greater surface brightness than the northern jet. This may indicate that we are seeing the jets of RG5 in projection, with the southern jet in the foreground (i.e. inclined somewhat towards us) and the northern jet in the background (i.e. inclined somewhat away from us). 

Overall, each jet extends some $\sim1.4$~arcmin from the core (a projected distance of $\sim95$~kpc) before blending into the larger-scale emission of what appears to be a radio lobe. Viewed in projection, we cannot determine to what extent there may be substructure in this lobe. From Figure~\ref{fig:RG5_Overlay}, the lobe exhibits a clear `kink' in its morphology: some 23~arcsec (25~kpc) beyond where the jet feeds into the larger structure, the lobe abruptly bends to the north before hooking to the east once more. The cause of this sharp change in morphology is unknown; this region lies beyond the region where our \xmm{} mosaic has sufficient signal-to-noise to probe the thermal properties of the ICM. Similarly, in the optical analysis of \cite{Dehghan2017}, there is no obvious structure (which might trace a matter overdensity) against which the expansion of this lobe might be hindered.

\subsubsection{Spectral properties}
The spectral index map of RG5 is presented in the right panel of Figure~\ref{fig:RG5_Overlay}. The core shows a typical active galaxy spectral index of $\alpha = -0.62 \pm 0.09$. While some fluctuations are visible, the jet-like structures generally show a fairly steep spectral index around $\alpha \simeq -1.3$ to $-1.5$ (with a typical uncertainty around $0.1$); there is no clear evidence of steepening along the length of the jets. At the presumed interface where the southern jet bends to the north and feeds into the larger lobe, the spectral index is remarkably uniform with a typical value of $\langle \alpha \rangle = -1.65 \pm 0.10$. Finally, in the lobe itself we observe a marked steepening to a median value of $\langle \alpha \rangle = -1.75 \pm 0.10$; in this region, the spectral index also appears to show increased fluctuations, perhaps indicating the mixing of different populations of electrons and/or a turbulent plasma. In places, the spectral index in the more extended lobe reaches values as steep as $\alpha = -2.2$, suggesting a particularly aged electron population.

\section{Conclusions}\label{sec:conclusions}
In this paper, we have presented new, deep multi-wavelength observations of the complex merging galaxy cluster Abell~3266. These observations were performed with the Australian Square Kilometre Pathfinder (ASKAP) at 943~MHz as an Early Science project for the Evolutionary Map of the Universe (EMU) survey, and combined with new dedicated observations from the Australia Telescope Compact Array (ATCA) at 2.1~GHz and re-analysis of ancillary \xmm{} data to perform a highly sensitive multi-wavelength study of this intriguing system.

Our observations have confirmed the presence of multiple sources of diffuse radio emission, including (i) a `wrong-way' radio relic, (ii) an ultra-steep spectrum fossil and (iii) an extended central diffuse `ridge' which we are as yet unable to classify. This `ridge' forms the brightest component of a larger-scale radio halo, which is revealed for the first time in our spatially-filtered ASKAP map at 943~MHz. We have performed a detailed investigation into the physical implications of these sources. Our main findings for each of the diffuse sources are as follows:

\begin{itemize}
    \setlength\itemsep{1em}
    \item \textbf{The `wrong-way' relic} (Source D1) lies to the far south-east of the cluster centre, and is coincident with an X-ray shock detected by eROSITA. The relic has a physical size of around 579~kpc at 943~MHz, and shows a clear spectral gradient trailing toward the cluster centre. The concave nature of the relic appears contradictory, but likely reflects the complex merger history of this cluster. However, the spectral energy distribution shows an unambiguous spectral steepening, which strongly rejects the standard scenario of diffusive shock acceleration from the thermal pool. The extreme steepening seen in the spectrum is exceptional for a radio relic and is not trivially explained by current radio relic formation scenarios. As such, further modelling is strongly motivated, but far beyond the scope of this paper. We measure a total radio power of $P_{\rm 1.4~GHz} = (2.38 \pm 0.10) \times 10^{23}$~W~Hz$^{-1}$ for the `wrong-way' relic in Abell~3266.
    
    \item \textbf{The ultra-steep spectrum fossil} (Source D2) lies to the north-west of the cluster centre. This source, first reported by the Phase II MWA, is resolved here for the first time with ASKAP; it is undetected by the ATCA at 2.1~GHz. Using our ASKAP data in conjunction with publicly-available MeerKAT data, we find clear evidence of a highly-curved spectrum. The radio power of this fossil estimated from our data is $P_{\rm 1.4~GHz} = (1.36 \pm 0.13) \times 10^{22}$~W~Hz$^{-1}$, although we note the true radio power may be lower. We identify a confirmed cluster-member galaxy as the likely original host of this source, suggesting that it is a remnant radio galaxy. No core is detected at either 943~MHz or 2.1~GHz, although the upper limit to the core prominence does not provide a tight constraint on the remnant nature.
    
    \item \textbf{The diffuse central `ridge'} (Source D3) is detected by ASKAP at 943~MHz but not by the ATCA at 2.1~GHz, implying an ultra-steep spectral index ($\alpha \lesssim -2.54$). We estimate an upper limit to the radio power of $P_{\rm 1.4~GHz} = (2.04 \pm 0.32) \times 10^{22}$~W~Hz$^{-1}$. The linear extent is around 240~kpc. While we conclude that it is not a classical radio halo like those hosted by many merging clusters, its true nature is puzzling. One possibility is that it is a mini-halo candidate, in which case it would exhibit a radio power nearly two orders of magnitude below the bulk of the known population. However, Abell~3266 shows no evidence of a cool core, which would be highly unusual (but not unprecedented) for a mini-halo. Alternatively, it could represent fossil plasma from cluster-member radio galaxies that has been gently re-accelerated by the cluster merger. This steep-spectrum plasma follows the edge of the low-entropy spine seen in our \xmm{} mosaic, suggesting it is bounded by this region. 
    
    \item We find low surface brightness diffuse synchrotron emission extending for a total size of up to 1.09~Mpc, confirming the presence of a \textbf{large-scale radio halo} after applying a multi-resolution spatial filter to our ASKAP map at 943~MHz. We perform a point-to-point analysis at 48~arcsec resolution, finding that the radio and X-ray surface brightness are strongly correlated, with a linear slope of $b = 0.99 ^{+0.07}_{-0.07}$ reflecting the behaviour in the $I_{\rm{R}}/I_{\rm{X}}$ plane. From our point-to-point analysis, we also find tentative evidence of substructure, suggesting we are viewing a multi-component radio halo or perhaps a mini-halo plus halo system. The brightest region of this radio halo is the `ridge' --- Source D3. The `ridge' shows a super-linear slope in the $I_{\rm{R}}/I_{\rm{X}}$ plane, while the larger-scale `halo' emission shows a clearly sub-linear slope. We measure an integrated flux density of $S_{\rm{943~MHz}}^{\rm{min}} = 42.2 \pm 4.3$~mJy and $S_{\rm{943~MHz}}^{\rm{max}} = 67.8 \pm 6.2$~mJy, corresponding to a 1.4~GHz radio luminosity of $P_{\rm 1.4~GHz}^{\rm min} = (1.94 \pm 0.20) \times 10^{23}$~W~Hz$^{-1}$ and $P_{\rm 1.4~GHz}^{\rm max} = (3.11 \pm 0.29) \times 10^{23}$~W~Hz$^{-1}$. These luminosity values place the radio halo in Abell~3266 a factor $4.4\times$ to $7.1\times$ below established correlations between cluster mass and radio halo power.
\end{itemize}

Additionally, our observations also allow us to perform a detailed resolved spectral study of three of the more spectacular resolved radio galaxies hosted by Abell~3266. These include the well known wide-angle tail radio galaxy and a complex ribbed radio galaxy recently reported by MeerKAT observations, as well as a less well-studied example to the north-east of the cluster. The spectral properties of these sources likely trace the rich dynamical history of the substructures within Abell~3266. However, further work is required to fully unpack the history of Abell~3266 and its constituent radio galaxies, and the answers to a number of questions remain elusive.

Finally, our results have demonstrated the depth and breadth of the detailed physics of cluster non-thermal phenomena and radio galaxies that can be extracted when applying next-generation techniques to exquisite multi-wavelength data from the SKA Pathfinder and Precursor instruments. The coming dawn of the SKA era promises a revolution in our understanding of these rich and complex environments.

\section*{Acknowledgements}
The Australia Telescope Compact Array is part of the Australia Telescope National Facility (\url{https://ror.org/05qajvd42}) which is funded by the Australian Government for operation as a National Facility managed by CSIRO. We acknowledge the Gomeroi people as the traditional owners of the Observatory site.

The Australian SKA Pathfinder is part of the Australia Telescope National Facility (\url{https://ror.org/05qajvd42}) which is managed by CSIRO. Operation of ASKAP is funded by the Australian Government with support from the National Collaborative Research Infrastructure Strategy. ASKAP uses the resources of the Pawsey Supercomputing Centre. Establishment of ASKAP, the Murchison Radio-astronomy Observatory and the Pawsey Supercomputing Centre are initiatives of the Australian Government, with support from the Government of Western Australia and the Science and Industry Endowment Fund. We acknowledge the Wajarri Yamatji people as the traditional owners of the Observatory site.

The MeerKAT Galaxy Cluster Legacy Survey (MGCLS) data products were provided by the South African Radio Astronomy Observatory and the MGCLS team and were derived from observations with the MeerKAT radio telescope. The MeerKAT telescope is operated by the South African Radio Astronomy Observatory, which is a facility of the National Research Foundation, an agency of the Department of Science and Innovation.

We thank our anonymous referee for their time and their positive and constructive feedback, which has improved our paper. CJR, EB and A.~Bonafede acknowledge financial support from the ERC Starting Grant `DRANOEL', number 714245. A.~Botteon acknowledges support from the VIDI research programme with project number 639.042.729, which is financed by the Netherlands Organisation for Scientific Research (NWO). KR and MB acknowledge financial support from the ERC Starting Grant `MAGCOW', no. 714196. JMD acknowledges the support of projects PGC2018-101814-B-100 and Mar\'ia de Maeztu, ref. MDM-2017-0765. This research was supported by the Australian Research Council Centre of Excellence for All Sky Astrophysics in 3 Dimensions (ASTRO 3D), through project number CE170100013. LDM is supported by the ERC-StG `ClustersXCosmo' grant agreement 716762.

This project used public archival data from the Dark Energy Survey (DES). Funding for the DES Projects has been provided by the U.S. Department of Energy, the U.S. National Science Foundation, the Ministry of Science and Education of Spain, the Science and Technology Facilities Council of the United Kingdom, the Higher Education Funding Council for England, the National Center for Supercomputing Applications at the University of Illinois at Urbana-Champaign, the Kavli Institute of Cosmological Physics at the University of Chicago, the Center for Cosmology and Astro-Particle Physics at the Ohio State University, the Mitchell Institute for Fundamental Physics and Astronomy at Texas A\&M University, Financiadora de Estudos e Projetos, Funda{\c c}{\~a}o Carlos Chagas Filho de Amparo {\`a} Pesquisa do Estado do Rio de Janeiro, Conselho Nacional de Desenvolvimento Cient{\'i}fico e Tecnol{\'o}gico and the Minist{\'e}rio da Ci{\^e}ncia, Tecnologia e Inova{\c c}{\~a}o, the Deutsche Forschungsgemeinschaft, and the Collaborating Institutions in the Dark Energy Survey.

The Collaborating Institutions are Argonne National Laboratory, the University of California at Santa Cruz, the University of Cambridge, Centro de Investigaciones Energ{\'e}ticas, Medioambientales y Tecnol{\'o}gicas-Madrid, the University of Chicago, University College London, the DES-Brazil Consortium, the University of Edinburgh, the Eidgen{\"o}ssische Technische Hochschule (ETH) Z{\"u}rich,  Fermi National Accelerator Laboratory, the University of Illinois at Urbana-Champaign, the Institut de Ci{\`e}ncies de l'Espai (IEEC/CSIC), the Institut de F{\'i}sica d'Altes Energies, Lawrence Berkeley National Laboratory, the Ludwig-Maximilians Universit{\"a}t M{\"u}nchen and the associated Excellence Cluster Universe, the University of Michigan, the National Optical Astronomy Observatory, the University of Nottingham, The Ohio State University, the OzDES Membership Consortium, the University of Pennsylvania, the University of Portsmouth, SLAC National Accelerator Laboratory, Stanford University, the University of Sussex, and Texas A\&M University.

Based in part on observations at Cerro Tololo Inter-American Observatory, National Optical Astronomy Observatory, which is operated by the Association of Universities for Research in Astronomy (AURA) under a cooperative agreement with the National Science Foundation.

Finally, several python packages were used extensively during this project but are not explicitly mentioned elsewhere in this paper. We wish to acknowledge the developers of the following packages: \textsc{aplpy} \citep{Robitaille2012}, \textsc{astropy} \citep{Astropy2013}, \textsc{cmasher} \citep{vanderVelden2020}, \textsc{colorcet} \citep{Kovesi2015}, \textsc{matplotlib} \citep{Hunter2007}, \textsc{numpy} \citep{Numpy2011} and \textsc{scipy} \citep{Jones2001}.

\section*{Data Availability}
The images underlying this article will be shared on reasonable request to the corresponding author. Raw visibilities from the ATCA are available via the Australia Telescope Online Archive (ATOA; \url{https://atoa.atnf.csiro.au}). Visibilities and pipeline-processed ASKAP images are available from CASDA (\url{https://research.csiro.au/casda/}). The \xmm{} data used in this paper are available through the Science Data Archive (\url{https://www.cosmos.esa.int/web/xmm-newton/xsa}).

\bibliographystyle{mnras}
\bibliography{Abell3266_ASKAP-EMU}

\section*{Supplementary}

\textbf{Figure A1.} shows ATCA images of Abell 3266 produced using standard techniques in \textsc{miriad} (\emph{left panel}) and using third-generation calibration techniques using \textsc{ddfacet} and \textsc{killms} (\emph{right panel}). Images are shown at 6-arcsec resolution and on matching colour scales, ranging between $-3\sigma$ and $200\sigma$ , where $\sigma = 14.5 \, \upmu$Jy beam$^{-1}$ is the rms noise in our \textsc{ddfacet} image. \\

\noindent
\textbf{Figure A2.} shows ASKAP images of Abell 3266 produced by \textsc{askapsoft} (\emph{left panel}) and with further post-processing using using \textsc{ddfacet} and \textsc{killms} (\emph{right panel}). Images are shown at $15.5 \times 13.7$ arcsec $\times$ arcsec resolution and on matching colour scales, ranging between $-3\sigma$ and $200\sigma$, where $\sigma = 36.7\, \upmu$Jy beam$^{-1}$ is the rms noise in our \textsc{ddfacet} image.

\section*{Affiliations}
\footnotesize{\textit{
$^1$ Dipartimento di Fisica e Astronomia, Universit\`a degli Studi di Bologna, via P. Gobetti 93/2, 40129 Bologna, Italy \\ 
$^2$ INAF -- Istituto di Radioastronomia, via P. Gobetti 101, 40129 Bologna, Italy \\ 
$^3$ CSIRO Space \& Astronomy, PO Box 1130, Bentley, WA 6102, Australia \\ 
$^4$ ICRAR, The University of Western Australia, 35 Stirling Hw, 6009 Crawley, Australia \\ 
$^5$ International Centre for Radio Astronomy Research, Curtin University, Bentley, WA 6102, Australia \\ 
$^6$ The University of Melbourne, School of Physics, Parkville, VIC 3010, Australia \\ 
$^7$ ARC Centre of Excellence for All Sky Astrophysics in 3 Dimensions (ASTRO 3D) \\ 
$^8$ CSIRO Space \& Astronomy, PO Box 76, Epping, NSW 1710, Australia \\ 
$^9$ Leiden Observatory, Leiden University, P.O. Box 9513, 2300 RA Leiden, The Netherlands \\
$^{10}$ Minnesota Institute for Astrophysics, University of Minnesota, 116 Church St. SE, Minneapolis, MN 55455, USA \\
$^{11}$ Th\"{u}ringer Landessternwarte, Sternwarte 5, 07778 Tautenburg, Germany \\
$^{12}$ Department of Astronomy, University of Geneva, Ch. d'Ecogia 16, CH-1290 Versoix, Switzerland \\
$^{13}$ GEPI \& USN, Observatoire de Paris, CNRS, Universite Paris
Diderot, 5 place Jules Janssen, 92190 Meudon, France \\
$^{14}$ Centre for Radio Astronomy Techniques and Technologies, Department of Physics and Electronics, Rhodes University, Grahamstown 6140, South Africa \\
$^{15}$ The Inter-University Institute for Data Intensive Astronomy (IDIA), Department of Astronomy, University of Cape Town, Private Bag X3, Rondebosch, 7701, South Africa \\
$^{16}$ School of Science, Western Sydney University, Locked Bag 1797, Penrith, NSW 2751, Australia \\
$^{17}$ Instituto de Fisica de Cantabria, CSIC-Universidad de Cantabria, E-39005 Santander, Spain \\ 
$^{18}$ Astronomy Unit, Department of Physics, University of Trieste, via Tiepolo 11, Trieste 34131, Italy \\
$^{19}$ INAF -- Osservatorio Astronomico di Trieste, via Tiepolo 11, Trieste 34131, Italy \\
$^{20}$ IFPU -- Institute for Fundamental Physics of the Universe, Via Beirut 2, 34014 Trieste, Italy \\
$^{21}$ Australian Astronomical Optics, Macquarie University, 105 Delhi Rd, North Ryde, NSW 2113, Australia \\
$^{22}$ Curtin Institute for Computation, Curtin University, GPO Box U1987, Perth, WA 6845, Australia \\ 
$^{23}$ Argelander-Institut f\"{u}r Astronomie, Universit\"{a}t Bonn, Auf dem H\"{u}gel 71, 53121 Bonn, Germany
}}

\bsp	
\label{lastpage}
\end{document}